\documentclass[12pt]{iopart}
\usepackage[utf8]{inputenc} % allow utf-8 input
\usepackage[T1]{fontenc}    % use 8-bit T1 fonts
\usepackage{hyperref}       % hyperlinks
\usepackage{url}            % simple URL typesetting
\usepackage{booktabs}       % professional-quality tables
\usepackage{microtype}      % microtypography
\usepackage{xcolor}         % colors

\usepackage{graphicx}
\usepackage{longtable}
\usepackage{multirow}
\usepackage{lscape}
\usepackage{csquotes}
\usepackage{caption}
\usepackage{subcaption}
\usepackage{natbib}
\usepackage{array}
\begin{document}

% % YKM, 29/11/2023
\title{Explainable artificial intelligence approaches for brain-computer interfaces: a review and design space}

% HC, 28/11/2023
% \title{A review of explainable artificial intelligence approaches for brain-computer interfaces}

% \title{XAI4BCI: elucidating the design space of explainable AI for brain-computer interfaces}

\author{Param Rajpura$^\dag$, Hubert Cecotti$^\S$, Yogesh Kumar Meena$^\ddag$}
\address{ $^\dag$ Department of Cognitive and Brain Sciences, Indian Institute of Technology Gandhinagar, India}
\address{ $^\S$ Department of Computer Science, California State University, Fresno, CA, USA}
\address{ $^\ddag$ Department of Computer Science \& Engineering, and Department of Cognitive and Brain Sciences, Indian Institute of Technology Gandhinagar, Gujarat, India}

\begin{indented}
\item[] December 2023
\end{indented}

\begin{abstract} 

\textit{Objective.} This review paper provides an integrated perspective of Explainable Artificial Intelligence (XAI) techniques applied to Brain-Computer Interfaces (BCIs). BCIs use predictive models to interpret brain signals for various high-stake applications. However, achieving explainability in these complex models is challenging as it compromises accuracy. Trust in these models can be established by incorporating reasoning or causal relationships from domain experts. The field of XAI has emerged to address the need for explainability across various stakeholders, but there is a lack of an integrated perspective in XAI for BCI (XAI4BCI) literature. It is necessary to differentiate key concepts like explainability, interpretability, and understanding, often used interchangeably in this context, and formulate a comprehensive framework. \textit{Approach.} To understand the need of XAI for BCI, we pose six key research questions (RQs) for a systematic review and meta-analysis, encompassing its purposes, applications, usability, and technical feasibility. We employ the PRISMA methodology -- preferred reporting items for systematic reviews and meta-analyses to review (n=1246) and analyze (n=84) studies published in 2015 and onwards for key insights. \textit{Main results.} The results highlight that current research primarily focuses on interpretability for developers and researchers, aiming to justify outcomes and enhance model performance. We discuss the unique approaches, advantages, and limitations of XAI4BCI from the literature. We draw insights from philosophy, psychology, and social sciences. We propose a design space for XAI4BCI, considering the evolving need to visualize and investigate predictive model outcomes customised for various stakeholders in the BCI development and deployment lifecycle. \textit{Significance.} This paper is the first to focus solely on reviewing XAI4BCI research articles. This systematic review and meta-analysis findings with the proposed design space prompt important discussions on establishing standards for BCI explanations, highlighting current limitations, and guiding the future of XAI in BCI.
\end{abstract}

\maketitle
% \ioptwocol
%

\section{Introduction}
\label{sec:introduction}

The explanation is \enquote{the details or reasons that someone gives to make something clear or easy to understand}, as defined by the Cambridge Dictionary \citep{walter2008cambridge}. \citet{arrieta2020explainable} further extend this definition in the context of explainable artificial intelligence (XAI) as \blockquote{given an audience, an explainable Artificial Intelligence is one that produces details or reasons to make its functioning clear or easy to understand.} 

One aspect of this definition stresses the ease of understanding the different steps leading to a decision. XAI generating "good" explanations may be translated to "ease" in understanding. This suggests that AI systems are inherently capable of giving explanations. For example, the statement: \textit{Neuron that predicts an outcome is fired due to subsequent activations of $n$ neurons in the previous layers}, tends to provide an explanation about the prediction but is not understandable enough for the diverse audience. Explaining a decision to a data scientist is not the same as explaining a decision to a neuroscientist or a patient. Furthermore, there is also a difference between explaining the algorithm and explaining the steps leading to a decision, or the quality of the decision. In particular, the accuracy, f-score~\citep{sokolova2006beyond}, Cohen-kappa~\citep{warrens2015five}, and the area under the receiver operating characteristic curve are measurements that quantify the performance of giving the expected decision, but they do not provide information explaining why such a level of performance is reached.

These discussions hint at the difference in the terms "explainability" and "understandability", often interchangeably used in the AI community. However, philosophical conceptualisation is beyond the scope of the paper; there is a renewed interest in the concept of understanding~\citep{de2009scientific,strevens2013no,strevens2011depth,potochnik2016scientific,de2017understanding,khalifa2017understanding}. \citet{erasmus2021interpretability, erasmus2022interpretability} propose that the process of interpretation is necessary to convert a poor explanation to a good explanation that leads to ease of understanding. 

The general structure of interpretation \citep{erasmus2021interpretability} allows us to: 1) provide a clear perspective on the terminology often used interchangeably in the field: explanation, understanding, and interpretation; 2) establish a general architecture for designing systems that can generate understanding via explanations post interpretations, and 3) build a unified view where the AI system may serve multiple kinds of audiences expecting different explanations and agnostic of the prediction model.

Likewise, interpreting brain signals to classify behaviour, diagnose patterns, and predict intended actions is crucial in Brain-Computer Interface's (BCI) acceptability for a broad audience. Recent state-of-the-art algorithms lack the transparency \citep{tjoa2020survey,doshi2017towards,tonekaboni2019clinicians,ribeiro2016should}, leading to a trust gap, especially in high stake applications in the medical and health domain. The lack of transparency or black-box nature can be attributed to the sizeable parametric space created by millions of parameters in the neural network model~\citep{castelvecchi2016can}. The notion of a trade-off between accuracy and explainability magnifies the motivation to address the trust gap while maintaining the accuracy of the AI systems irrespective of their field of application~\citep{london2019artificial}.

\begin{figure*}[t!]
\begin{center}
\includegraphics[width=0.9\textwidth]{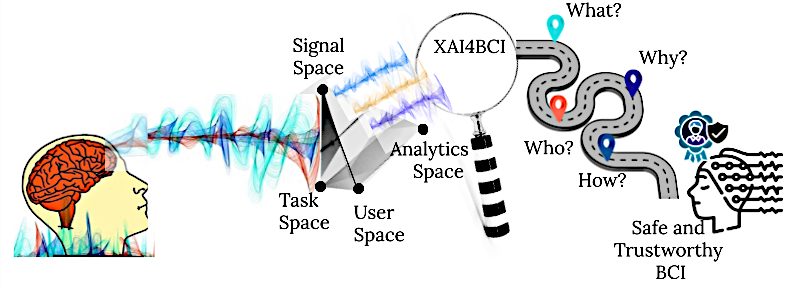}
\caption{Explainable AI as a lens for developing safe and trustworthy BCIs.}
\label{fig:XAIoverview}
\end{center}
\end{figure*}

% To revise
% The concept of Brain-Computer Interface (BCI) is not easy to convey to the public as it deals with the analysis of brain responses, which are directly related to the mind. The direction of the signals, from the brain to the machine, and what is analyzed is often not clearly explained to the general public. For instance, detecting the amplitude of event-related potential components at the single-trial level is not mind reading, and it should be clearly explained to the people using BCI. 

BCI encompasses transmitting signals from the brain to the machine and analyzing these brain responses. This process often needs to be more adequately explained to a broader audience since it leads to a direct association with the mind. For example, discerning the amplitude of event-related potential components at the single-trial level does not constitute mind reading. It is imperative to articulate this distinction clearly to individuals utilizing BCI.

\begin{figure*}[t!]
\begin{center}
\includegraphics[width=0.95\textwidth]{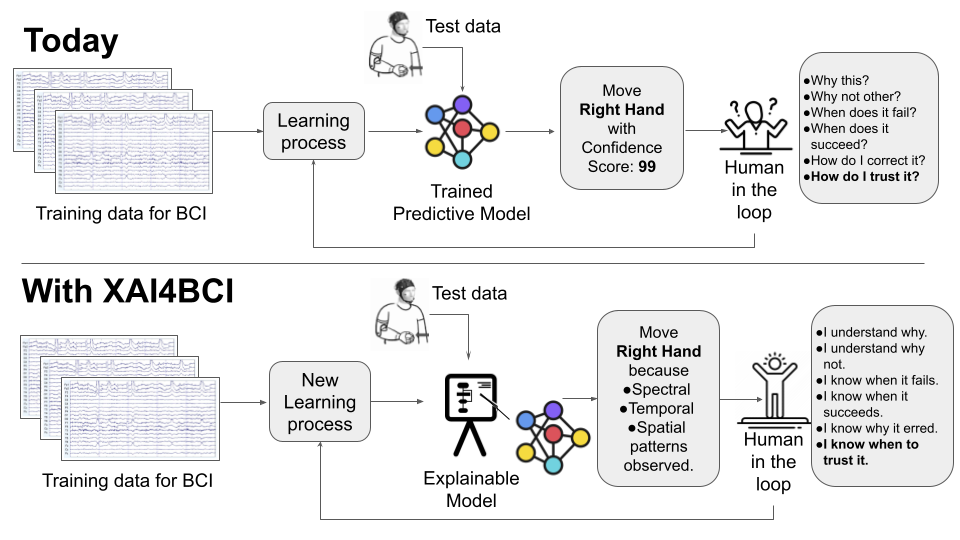}
\caption{Purpose of Explainable AI for Brain-Computer Interfaces to assist the human-in-the-loop. Without XAI, BCI applications using complex predictive models often turn out to be black-box approaches, less transparent, and unreliable. }\label{fig:XAIpurpose}
\end{center}
\end{figure*}

\autoref{fig:XAIoverview} illustrates XAI's larger perspective and position to address the trust gap and the necessity of evolving the existing design space for BCI. Through the metaphorical prism with four parameters of the design space, insights are derived about brain activity but are insufficient. The four parameters are: $1)$ the \textit{Signal space} defines the means of capturing brain activity, $2)$ the \textit{Task space} defines the cognitive activity, $3)$ the \textit{User space} represents the individuality in signals, and $4)$ the \textit{Analytics space} represents the choice of signal processing techniques for analysis. XAI was introduced as an essential research goal to build models that could explain their predictions and enable humans to understand and trust without compromising accuracy \citep{gunning2017explainable}. It aims to bridge the trust gap arising from black box models. Thereafter, various stakeholders (developers, researchers, and policymakers) constantly discuss the definition of explainability and its necessity, widening the scope and approach towards XAI. By incorporating XAI as a critical lens, a roadmap toward safe and trustworthy BCIs can be defined. The purpose of XAI for BCI can be translated to assisting the human-in-the-loop, either as a user or an indirect stakeholder (see \autoref{fig:XAIpurpose}). 

Our research further studies explainability, interpretability, and understanding in BCI, and it adapts the general explanation-interpretation-understanding structure to discuss XAI for BCI~\citep{erasmus2021interpretability}.
The primary contributions of this topical review paper can be outlined as follows: 

\begin{enumerate}
    \item A comprehensive investigation aimed at collecting and analyzing the literature regarding the methods and strategies employed to enhance trust in predictive models for BCI through the incorporation of explainability and interpretability.
    \item Address the six research questions (RQs) via a systematic review and meta-analysis of current BCI literature. \textbf{RQ1}--Why does one need explanations for BCI performance?; \textbf{RQ2}-- Who are the stakeholders needing an explanation for BCI?; \textbf{RQ3}--What are the application domains of BCI that can be facilitated by explanations?;\textbf{ RQ4}--How are the explanations for BCI generated and communicated across the stakeholders?; \textbf{RQ5}--What is the design space for building transparent and trustworthy BCI?; and \textbf{RQ6}--How can this design space practically facilitate building a BCI?
    \item Construct a broader design space imbibing XAI4BCI with its four key variables (i.e., What, Why, Who, How) and their corresponding categories. This comprehensive perspective facilitates the design of BCI systems that incorporate XAI. 
\end{enumerate}

The remainder of the paper is organized as follows. Section~\ref{sec:review} discusses the review methodology. Section~\ref{sec:ecosystem} further investigates RQ1 and RQ2 by analysing the motivations of published studies applying XAI to BCI, answering RQ3 by categorizing the applications and datasets addressed in the literature. By enlisting the XAI techniques and the interfaces used for explanations in studies, we scrutinize RQ4. Section~\ref{sec:designspace} describes the design space and tackles RQ5 and RQ6. Finally, the impact and limitations of XAIBCI are discussed in Section~\ref{sec:discussion}.

\section{Review methodology}
\label{sec:review}

We conducted a systematic review to assimilate the answers to RQ1-4. In this section, we discuss in detail the review planning, execution, and insights from the literature adhering to the PRISMA 2020 protocol. The objectives, review protocol, and data collection procedure are reported.

\subsection{Identifying the need for a systematic review}
% \textbf{Identifying The Need For A Systematic Review:}

 Citing the trust gap addressed in the previous section, the knowledge of XAI methodologies applied to BCI is currently unorganised. There was no evidence of a systematic review conducted prior to this work in the investigated bibliographic databases. Therefore, this review is intended to compile and analyse the publications on the methods and strategies applied to bridge the trust gap of predictive models for BCI via explanations and investigate RQ1-4. This review aims to assist the researchers and stakeholders to understand the current state of the art and identify future research directions in XAI for BCI.

\subsection{Systematic review protocol}

We follow the PRISMA 2020 protocol for systematic review and meta-analysis~\citep{page2021prisma,prisma2021}. The inclusion and exclusion criteria are as follows: 

\begin{itemize}
    \item{Keywords}: (XAI OR explainable OR interpretable) AND (BCI OR Brain Computer Interface).
    \item{Databases}: Scopus, Google Scholar, Springer, IEEE, JMIR, AMIA, ScienceDirect, ACM
    \item{Last search conducted}: 20th March 2023
    \item{Exclusion Criteria}: Study not written in English. Study that did not include a method for interpretation or explanation of the model's outcomes. Studies published before 1st Jan 2015.
\end{itemize}

\begin{figure}[t!]
\begin{center}
\includegraphics[width=\textwidth]{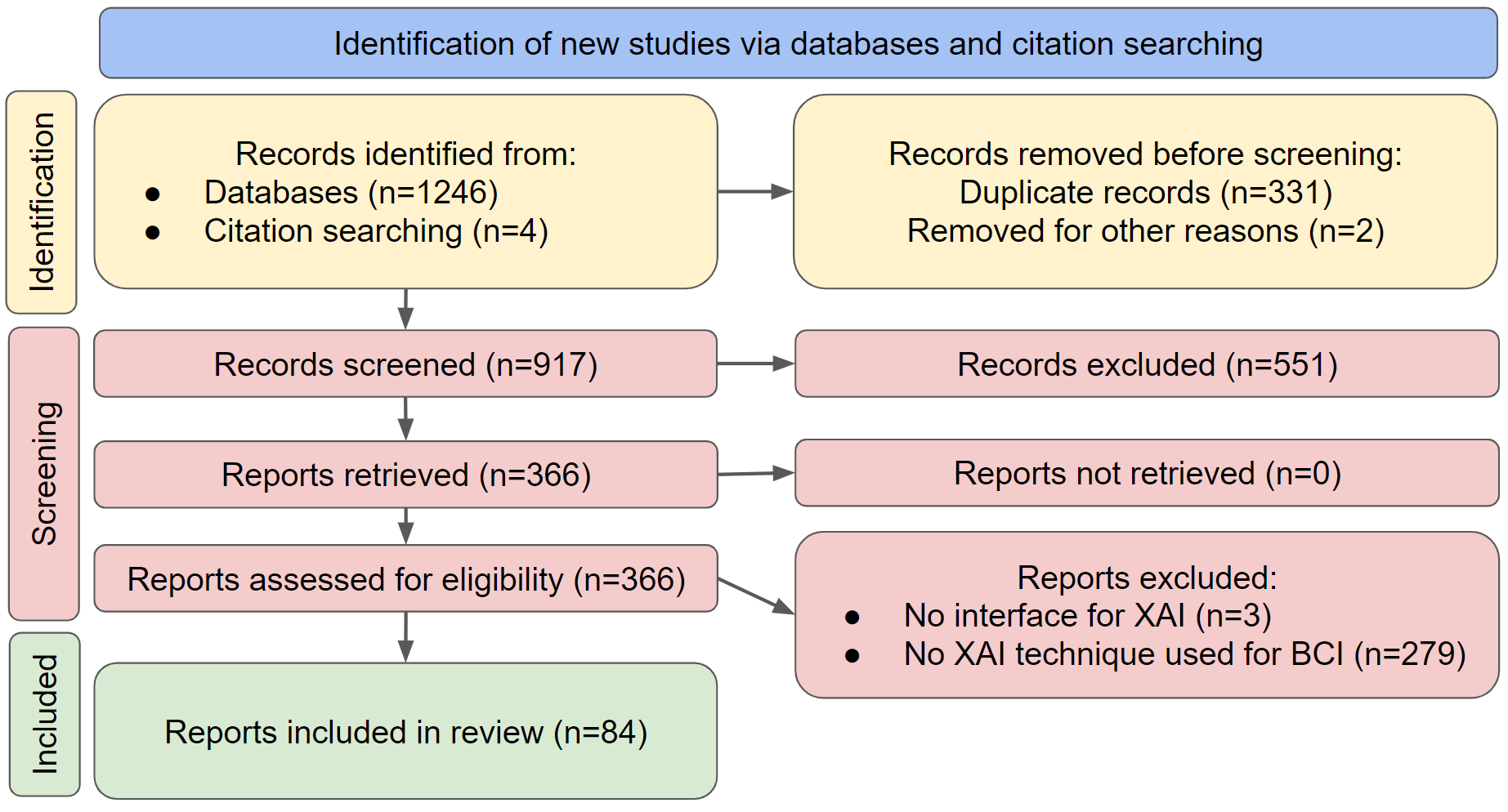}
\caption{ The step-wise flow diagram for identifying the selected articles as recommended by PRISMA 2020 protocols.}
\label{fig:PRISMA}
\end{center}
\end{figure}

Adding the keyword \textit{review, meta-analysis, survey} to the query had no results, suggesting no prior research was conducted on XAI for BCI. The stepwise flow diagram for identifying the sources for review is depicted in \autoref{fig:PRISMA}. The search retrieved 1252 research studies. Other alternate sources were not considered since they primarily indexed data from the mentioned sources. For a broader search, we included words like interpretable and explainable to cover literature that did not explicitly mention XAI. Moreover, we observed that Google Scholar results contained a large number of unselected studies due to its search within the text and citations of the research study. After removing the duplicate entries and studies published in languages other than English, 917 records were screened. After reviewing the titles or abstracts, 551 records were excluded. Full reports from the remaining 366 records were accessed to verify if they applied or proposed an explanation of the model outcomes and also represented the findings in any form. We excluded studies where keywords like interpretable and explainable features were used but did not describe methods for explanations or interpreting the features~\citep{yin2020locally}. Therefore, 84 articles, including four articles identified from the citation search were selected for a thorough review.

\subsection{Data collection and synthesis}

 All the selected studies were scrutinized by a four-point questionnaire to identify the application domain, XAI technique, their objective for providing explanations on outcomes, and representations used for explanations, fulfilling the objectives for the systematic review. The process and the collected data were reviewed by the co-author to ensure the reporting was unbiased and quoted direct answers from manuscripts. For the analysis stage, in order to systematically synthesize the results, reviews on definitions and taxonomy involving XAI and BCI \citep{dhanorkar2021needs,miller2019explanation,arrieta2020explainable,saeed2023explainable,ding2022explainability,moore2003real,moore2010applications,mak2009clinical} were considered. To analyse each research question(RQ), we use the number of published studies satisfying each category as an effective measure to indicate the trend. It is essential to note that the objective of each research study (RQ1) was extracted from the full text and classified into categories based on the author's discretion. Similarly, the XAI methods (RQ4) were grouped into broader categories based on the underlying concepts after referring to the relevant literature~\citep{speith2022review,adadi2018peeking,angelov2021explainable,arya2019one,arrieta2020explainable}. The details on the categorisation are discussed while analysing the results in the next section. 
% \begin{figure}[t!]
%     \centering
%      \begin{subfigure}[b]{0.4\textwidth}
%          \centering
%          \includegraphics[trim={1.8cm 0cm 2cm 0cm},width=\linewidth]{year_trend.png}
%          \caption{}
%          \label{fig:yeartrend}
%      \end{subfigure}
%     \hfill
%     \centering
%      \begin{subfigure}[b]{0.55\textwidth}
%          \centering
%          \includegraphics[trim={1.8cm 2cm 2cm 0cm},width=\linewidth]{publishedSources.png}
%          \caption{}
%          \label{fig:pubSources}
%      \end{subfigure}
%         \caption{a) The number of published studies grew three times from 2021 to 2022 and has already covered a quarter of the estimates until March 2023. b) The sources of the studies referred to in the review.}
%         \label{fig:pubstats}
% \end{figure}

\begin{figure}[t!]
\begin{center}
\includegraphics[trim={1.8cm 0cm 2cm 0cm},width=\linewidth]{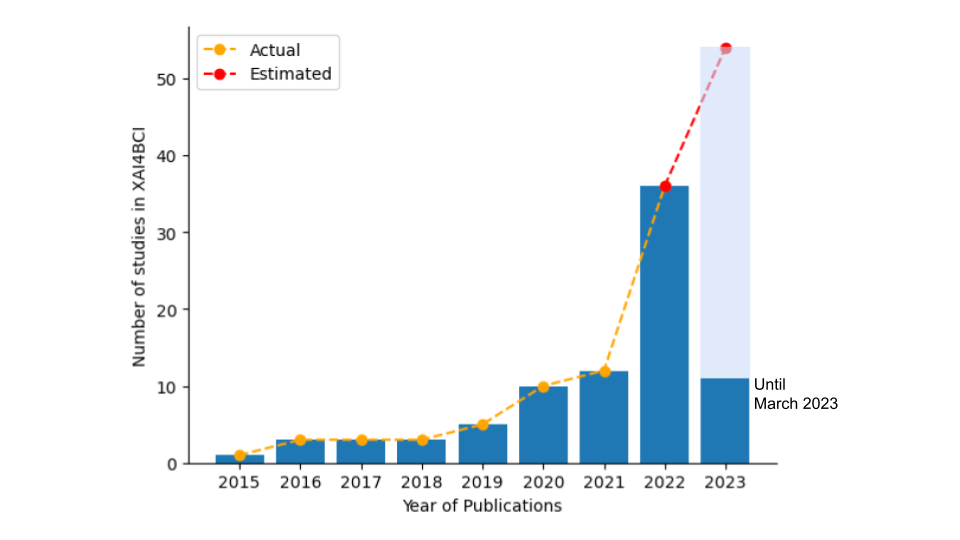}
\caption{The number of published studies grew three times from 2021 to 2022 and has already covered a quarter of the estimates until March 2023.}\label{fig:yeartrend}
\end{center}
\end{figure}

\begin{figure}[t!]
\begin{center}
\includegraphics[trim={1.8cm 1.8cm 2cm 0cm},width=\linewidth]{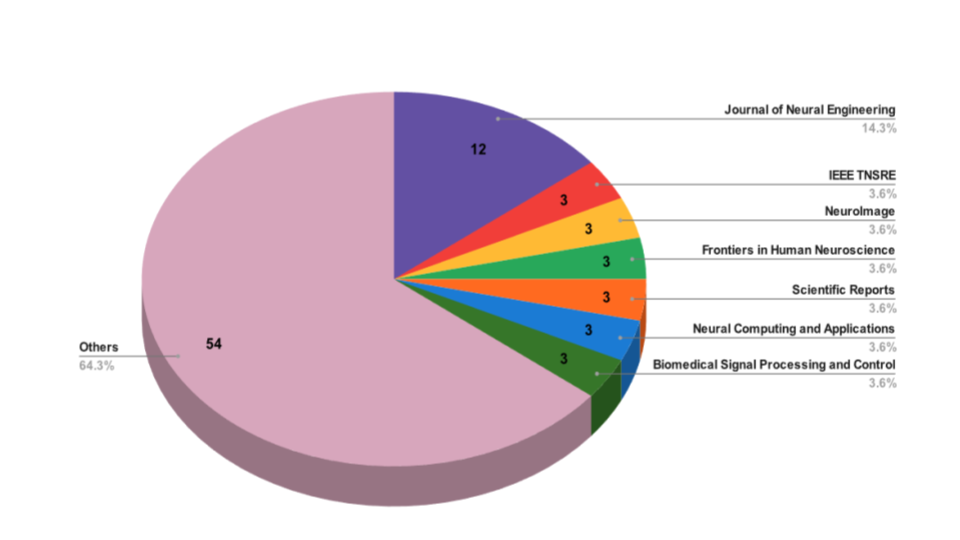}
\caption{The sources of the studies referred to in the review. The number on the pie chart indicates the number of studies cited in this review published at a given source. }\label{fig:pubSources}
\end{center}
\end{figure}

\section{Unveiling the BCI explanation ecosystem}
\label{sec:ecosystem}

The number of publications has increased over time, with exponential growth after 2018. It suggests that the predictive models have been successful at interpreting brain signals but need relevant explanations to be trustworthy. \autoref{fig:yeartrend} shows the growth rate and the estimations from the historical data by curve fitting using an exponential function. The evolving need to discuss future directions and roadmap is evident from the trend. \autoref{fig:pubSources} represents the top publication sources of the studies covered in this literature review. This section discusses the findings from the review and discusses the meta-perspective on each research question. 

\begin{figure}[t!]
\begin{center}
\includegraphics[width=0.7\linewidth]{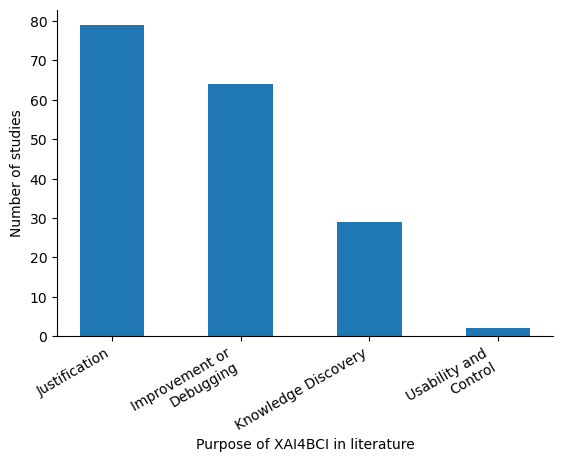}
\caption{Purpose for applied XAI for BCI studies.}\label{fig:why}
\end{center}
\end{figure}

\subsection{Demystifying the explanations in BCI (RQ1)}

The fundamental rationale for elucidating predictions through models in BCI lies in addressing the drivers behind explaining AI. These drivers will be helpful while conceptualising interpretation models that build understandable explanations over the raw explanations generated by core AI models. 
To analyse the purpose of explanations after the data collection phase, each study was tagged with its purpose by categorizing the keywords from their respective motivation for the study. The categorization was subjective since the choice of keywords was based on the authors' domain knowledge. \autoref{fig:why} illustrates that seeking justification and appropriate neurophysiological validation for the features learned by the respective models is the leading motivation to employ XAI for BCI, followed by its applications for model improvement and debugging.

\subsection{Stakeholder demands (RQ2)}
% \textbf{(RQ2) Who needs explanations in BCI?:} 

The definition of explanation hinted at the significance of the audience~\citet{arrieta2020explainable}. \citet{miller2019explanation} draws crucial insights from the social sciences on how people use different models or strategies to define, generate, select, present, and evaluate explanations. \autoref{fig:why} shows that researchers are the relevant stakeholders seeking justification, while it is more apt from a developer's perspective to use explanations in probing the models, improving, and debugging the model performance.

\begin{figure}[t!]
\begin{center}
\includegraphics[width=0.75\linewidth]{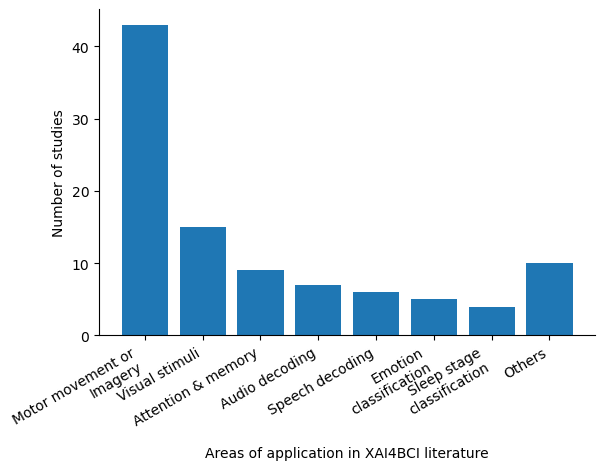}
\caption{Trend of the cognitive tasks that employed XAI techniques for BCI. Motor movement and imagery is one of the most studied domains in BCI with explainability.}\label{fig:what}
\end{center}
\end{figure}

\subsection{Enhancing BCI applications (RQ3)}
% \textbf{(RQ3) What applications in BCI need explanations?:} 

Applications of BCI need to be considered to understand what kind of explanations are required for them. By categorising the outcomes that are predicted, it is possible to map the "how" to "what", "who" and "why" for the explanations.

BCI applications have grown from assistive devices for communication and rehabilitation to neurofeedback and recreational technologies. While the research studies do not explicitly draw out specific target application domains, the underlying datasets can suggest the nature of tasks related to BCI applications. The tasks the BCI users performed for the research studies represent the trend in \autoref{fig:what}. A noteworthy interest in decoding motor imagery and movement for assistive technology and rehabilitation applications is evident with a rise in demand for public datasets~\cite{gwon2023review}.

The increasing trend for applications indicates that BCI also carries the promise of mass acceptance in low-risk domains like gaming and recreational technology. However, this highlights an opportunity to build common platforms to test the XAI techniques across different BCI datasets and cognitive tasks for benchmarking model performance and evaluating their explanations. Further, the interest in qualitative and quantitative evaluation of explanations\citep{markus2021role, nauta2022anecdotal} needs to be translated and discussed in the context of BCI. Encouraging the practice to benchmark novel models and techniques for performance and explainability across various domains\citep{kim2021deep,caywood2017gaussian,lekova2021fuzzy,qu2022eeg4home,du2022ienet} is a necessary step to design effective and generalisable tools for BCI.

\begin{figure}[t!]
\begin{center}
\includegraphics[width=\linewidth]{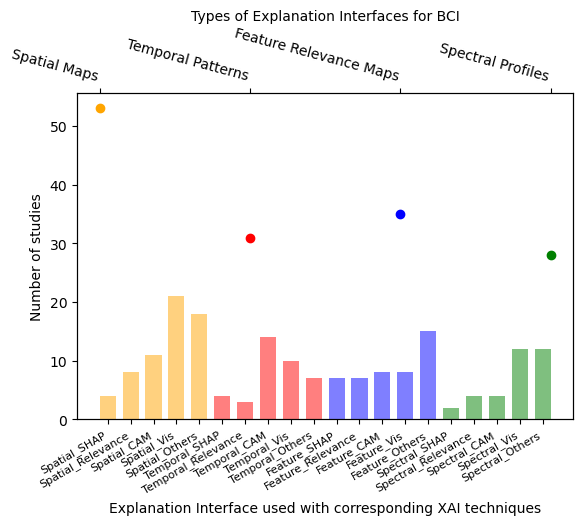}
\caption{Trend of the interfaces employed to communicate the explanations to stakeholders.}
\label{fig:how}
\end{center}
\end{figure}

\subsection{Communicating BCI insights (RQ4)}

Explanations can be easily understood by the stakeholders using appropriate interfaces for explanations. The literature review revealed the nature of interpretation methods and the commonly used interfaces in BCI and is visualised in \autoref{fig:how}. In the figure, the labels on the lower axis of the graph are a combination of the XAI technique and the corresponding interface used in the research study. The histograms highlight the trend of using visualisation of weights and attention maps for spatial maps as explanations.

The XAI techniques were grouped into $1)$ SHAP(Shapley Additive Explanations)\citep{lundberg2017unified} based methods including SHAP and DeepLIFT/DeepSHAP\citep{shrikumar2017learning}, $2)$ Relevance based methods including Layerwise Relevance Propagation\citep{bach2015pixel} and Integrated Gradients\citep{sundararajan2017axiomatic}, $3)$ Class Activation Map (CAM) based methods including Saliency maps\citep{simonyan2013deep}, GradCAM \citep{selvaraju2017grad} etc. and $4)$ Visualisation techniques using attention, learnt custom filters or weights. Other techniques like LIME\citep{ribeiro2016should}, Tree-based, t-Distributed Stochastic Neighbor Embedding (t-SNE)\citep{van2008visualizing}, Occlusion Sensitivity Analysis, Fuzzy rule interpretation, etc. were grouped into $5)$ Others.

Each study communicated the explanations using an interface that was populated by employing the respective XAI technique. Spatial maps that visually helped interpret the brain regions and the spatial patterns generated while performing the designated cognitive task can be said to reveal more insights to the stakeholders based on the analysis. \citet{kim2023designing} address the requirement of designing XAI interface for experts, initiating a necessary research discussion. \citet{aellen2021convolutional} use XAI to understand the trial-by-trial changes in features learnt by models to debug and improve the performance. Various studies \citep{du2022ienet,lawhern2018eegnet,cui2022eeg} demonstrate the application of multiple XAI techniques to cover all four aspects(spatial, spectral, temporal, and feature relevance maps) of explanation interfaces for BCI.
Various studies\citep{petrosyan2021compact,petrosyan2021decoding,petrosyan2022speech} approach interpretability by transformation of sensor signals into the source space in the model. \citep{ahn2012feasibility} reveal that sensor space and source space provide complementary information for building accurate BCIs. Building models by incorporating domain knowledge about source space is a distinctive approach for XAI4BCI to leverage.

The role of data-driven statistical explanations is evident. However, causal and counterfactual explanation and intervention\citep{karimi2021algorithmic} have an essential role to play, which remains unexplored. It is essential to explore further if non-causal (answering: what happened?) explanations are sufficient or causal (answering: why it happened?) explanations are necessary for the stakeholder to fulfill their respective motivation to seek an explanation. 

% We will add an interactive tool here for chord diagram 
\begin{figure}[t!]
\begin{center}
\includegraphics[width=\linewidth]{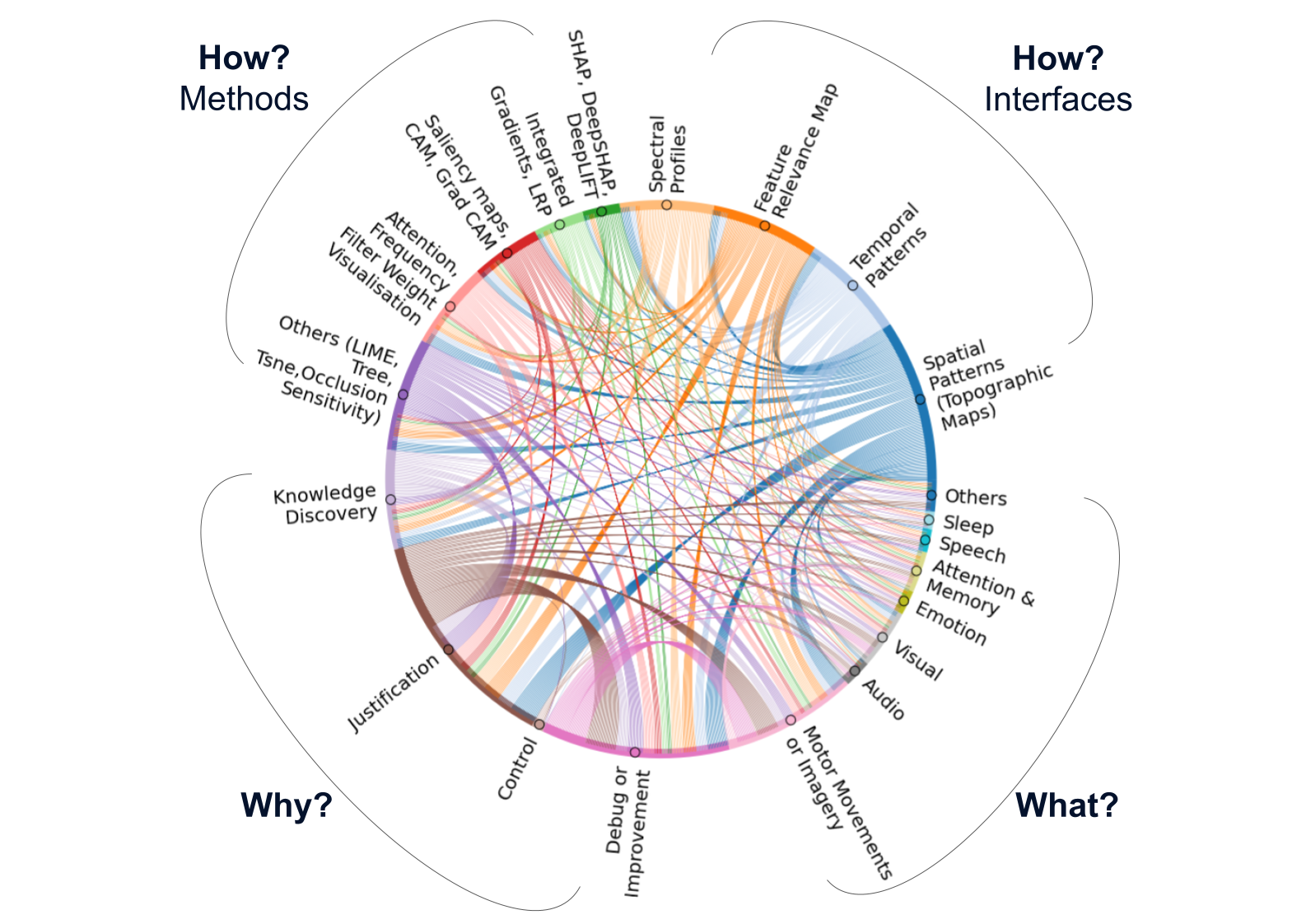}
\caption{Chord diagram relating all factors considered for the review. An open-source interactive version of the chord diagram enabling custom visualisations to better comprehend the data is shared on a StreamLit application (\href{https://xai4bci.streamlit.app/}{Link}).}
\label{fig:chord}
\end{center}
\end{figure}

\begin{figure*}[t!]
\begin{center}
\includegraphics[width=0.9\textwidth]{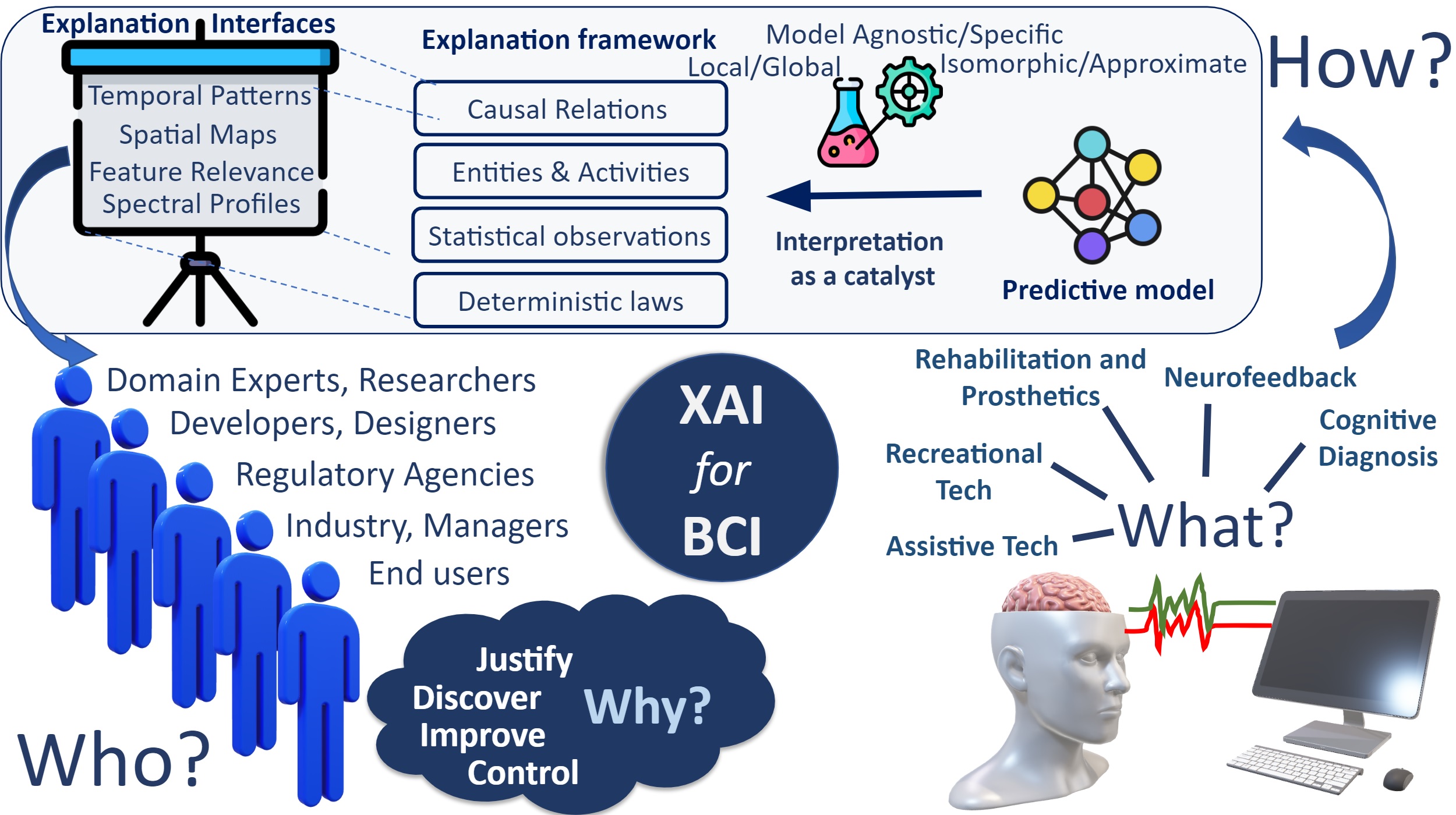}
\caption{XAI4BCI: Proposed Design Space for developing Explainable Brain-Computer Interfaces. The figure gives a meta-view of the constituent variables: Why?, Who?, What? and How? for the XAI4BCI design space. }\label{fig:overview}
\end{center}
\end{figure*}
 
\section{XAI4BCI design space}
\label{sec:designspace}

The combinatorial space formed by the primitives defines the design space ~\citep{stankiewicz2000concept}. We discussed those primitives for BCI in the previous section. Considering the evolution of design spaces as the technological growth, adding XAI to the BCI design space is an opportunity to take the next leap towards the adaptation of BCI with human-in-the-loop and align towards the goal of Human-centred AI~\citep{shneiderman2020human}. We assimilate the findings with a meta-view of the data from the systematic review, visualising all the factors and variables involved in the domain. \autoref{fig:chord} aids to visualise the design space variables and the categories that are correlated across the reviewed papers using a chord diagram. A chord diagram is a graphical method of displaying the relationships between data in a matrix. It is used to visualize the connections or relationships between a finite set of entities. The entities are represented as circular segments, and the connections between them are illustrated using ribbons or chords. The entities are represented as circles around the perimeter of the diagram, each circle corresponds to an individual entity. The chords are the arcs connecting pairs of entities. The width of the chord may be used to represent the strength or frequency of the relationship between the connected entities. The input is a matrix, where the rows and columns correspond to the entities, and the matrix entries represent the strength or frequency of the relationships between them. The color represents the different entities around the disk. The colors are chosen arbitrarily. An open-source interactive version of the chord diagram enabling custom visualisations to comprehend the data better is shared on a StreamLit application (\href{https://xai4bci.streamlit.app/}{Link}).
We extrapolate the variables to construct a broader design space: XAI4BCI illustrated in \autoref{fig:overview} as a solution to RQ5. 

The design space has four key variables, and their corresponding categories bring forth the comprehensive perspective to aid the design of any BCI system encompassing XAI.

\textbf{What?}: Categorising a BCI system to its application domain guides the decision on other variables and identifies their suitable categories. For example, BCI designed for recreational technologies may not need an in-depth explanation of the decisions for the user as a BCI for Cognitive Diagnosis.
BCI applications coinciding with the previous reviews \citep{mak2009clinical,moore2010applications,moore2003real} are categorised as:
\begin{itemize}
    \item \textit{Assistive Technologies}: for Communication, Mobility, Environment Control, and Augmented Cognition.
    \item \textit{Rehabilitation and Prosthetics}: to evaluate, and promote the function and mobility following amputation or brain injuries. 
    \item \textit{Cognitive Diagnosis and Treatment}: for Disease Detection, Neurofeedback and Training.
    \item \textit{Recreational Technologies}: for Gaming, E-Sport, Virtual Reality, and Generative Art.
\end{itemize}

\textbf{Why?}: The purpose of the explanation in BCI guides the interface and methods to generate explanations for the targeted user. \citet{adadi2018peeking,ding2022explainability} propose the motivations for explanation as follows: 

\begin{itemize}
    \item \textit{Explain to discover}: Explanations can be designed to discover new knowledge. For example, models predicting outcomes from large datasets used in BCI can help discover new knowledge about the underlying functions of the brain and regions involved in processing specific sensory stimuli or decision-making. 
    \item \textit{Explain to justify}: Stakeholders need justification of predictions generated by models, especially for high-stake outcomes, to establish trust and evaluate the reliability of AI systems. For example, BCI, used for early diagnosis of a brain disease, may need to highlight certain abnormalities in the brain signals to justify its decision to experts.  
    \item \textit{Explain to control}: BCI system users often require calibration to use them effectively. Explanations about the mechanisms underlying the assistive interfaces can help users in their effective control.
    \item \textit{Explain to improve}: AI systems are usually improved by active or passive feedback loops. Explanations that effectively highlight the relevant statistics and causal relationship to the outcome can help improve the system against bias and noise in the data.
\end{itemize}

\textbf{Who?}: Explanations need to consider the human-in-the-loop aspect at various stages of the BCI lifecycle. Extending the context: "Why?", BCI stakeholders may need explanations for one or multiple reasons to validate their will to consider a BCI. In the XAI4BCI design space, the following stakeholders drive the demands to interpret models and generate relevant explanations: 
\begin{itemize}
    \item \textit{End users}: individuals intended to be the direct consumers of the BCI including healthy individuals, patients, clients for service providers, etc. It is even more important for invasive BCI where patients must fully know all the risks and benefits in the short term and long term.
    \item \textit{Domain experts or Researchers}: the experts in the field of neuroscience, physiology, or domain experts in the application areas of BCI.  
    \item \textit{Industry, Managers, and Executives}: individuals involved indirectly in the conceptualisation, production, and commercialisation of BCI or related services.
    \item \textit{Regulatory agencies}: the individuals involved in policy-making and its enforcement, actively involved in the governance and ethical regulations for BCI. 
    \item \textit{Developers, Designers}: individuals who are directly involved in designing and engineering BCI systems.
\end{itemize}

\textbf{How?}: To generate explanations considering the aforementioned variables, an interpretation is necessary for understanding the origins of the decisions. As per the general model of explanation, interpretation, and understanding \citep{erasmus2021interpretability}, it is crucial to identify the types of explanations available in BCI. Hence, a satisfactory explanation may fall into one or a combination of the following categories:
\begin{itemize}

    \item \textit{Law-like statements}: deterministic laws of physics and chemistry underneath the cognitive processes
    \item \textit{Statistical laws}: probabilistic statements confirmed by data or data-driven models from brain signals,  
    \item \textit{Causal relationships}: causal process and interactions driving the phenomenon derived from causal inferences using AI or expert knowledge. Often information from brain lesion studies and brain stimulation provide causal relationships for brain functions\citep{siddiqi2022causal}
    \item \textit{Entities and activities}: the entities and the corresponding activities evolve into mechanisms via explanations. Identified networks acting as entities activated in brain regions can be depicted as a mechanism.
\end{itemize}

Multiple models already have properties to facilitate their explanations. Linear models, e.g., linear regression or linear support vector machines, are inherently interpretable. The coefficients associated with each feature provide an understanding of the impact each feature has on the model's output.
Decision trees are tree-like structures where each node represents a decision based on a feature, and each leaf node represents the final decision or output. They are interpretable, and the paths through the tree can be traced to understand the decision-making process. Rule-based systems, e.g., expert systems, express decision logic in the form of "if-then" rules. These rules are human-readable and provide a transparent way to understand how the system makes decisions. While neural networks are often considered black boxes, there are techniques for making neural networks more interpretable such as attention mechanisms~\citep{niu2021review} that highlight specific parts of the input that the model focuses on, and layer-wise relevance propagation (LRP)~\citep{montavon2019layer} that attributes relevance to each input feature. The choice of an explainable AI technique for BCI depends on the type of brain responses to detect, the complexity of the model, and the desired level of interpretability, if it is necessary to go to the source level with the definition of the activated brain regions or not.

An explanation on the level of predictive model when not sufficient, needs an interpretation categorized as one or multiple among the following:  

\begin{itemize}
    \item \textit{Global or Local interpretations}: The interpretation may be specific to an instance of the possible input-output combinations. For example, explaining to justify a decision by a BCI-driven robot to take a right-side turn, is better understood by explaining the particular instance(local) than explaining the underlying rules of navigation(global) learned by the model. Hence there can be global or local interpretations.  
    \item \textit{Approximate or Isomorphic interpretations}: During the process of interpretation of complex explanations, approximation helps in understanding. For example, instead of explaining all the features that were relevant in diagnosing an abnormality in brain activity, approximating the complex explanation by providing prominent features may be simple and understandable for justification.
    \item  \textit{Model-specific or Model-agnostic interpretations}: Interpretation techniques can be very specific to the chosen model or model-agnostic. Recent techniques are often meant for the interpretation of weights learnt by a neural network. While few of the techniques are specific to the architecture, some are agnostic.
\end{itemize}

An interface is a crucial facilitator for explanations. For BCI, an apt explanation could focus on the following categories to generate a comprehensive perspective:
\begin{itemize}
    \item \textit{Spectral maps}: Brain oscillations are indicative of cognitive activity and facilitate unique perspectives to represent them.   
    \item \textit{Temporal patterns}: Specific temporal patterns have been identified to represent known stimuli or cognitive activity. When signals captured are phase-locked with the stimuli, responses can be compared across individuals and common patterns are identified. 
    \item \textit{Spatial maps}: Spatial maps facilitate visualisation in 2 or 3 dimensions across different modalities for measuring brain activity, represented either on the sensor level (location of the sensor) or source level (identified source brain region). 
    \item \textit{Feature relevance maps}: Considering the measured signals as features for the predictive models, relevance scores signify the role of particular features in decision-making. 
\end{itemize}

We defined the four axes of design space: Who, Why, What, and How keeping the explanations for human-in-the-loop as the pivot. Each design space variable, and its categories with the relevant references from the current literature are tabulated in \autoref{tab:DSWhy} (Why?), \autoref{tab:DSWhat} (What?), \autoref{tab:DSHowMethods}(How? : XAI Methods) and \autoref{tab:DSHowInterfaces}(How? : XAI Interfaces). 
% listed as follows:

% \onecolumn
% \begin{landscape}

\setlength\LTleft{0cm}\begin{longtable}{ll p{12cm}}
\caption{XAI4BCI Design space variable: \textit{Why?}, its categorisation and corresponding references from current literature}
\label{tab:DSWhy}\\
\hline
\multicolumn{1}{c}{\textbf{Categories}} & \multicolumn{1}{c}{\textbf{References}} \\ \hline
\endhead
% \multicolumn{1}{c}{} & 
% \rotatebox[origin=c]{90}{Justification} & \begin{tabular}[c]{{ @{} p{\dimexpr0.95\textwidth} @{} }}
\begin{tabular}[c]{@{}l@{}}Justification\end{tabular} & \begin{tabular}[c]{{ @{} p{\dimexpr0.95\textwidth} @{} }}
\cite{urdanetaexplainable,song2022eeg,dutt2022sleepxai,xie2022transformer,giudice2022visual,kim2023designing,nagarajan2022relevance,kim2023identification,shibu2022explainable,petrosyan2022speech,petrosyan2021decoding,kumar2022neurophysiologically,fu2022single,chen2022novel,lawhern2018eegnet,zygierewicz2022decoding,borra2022bayesian,zhao2019learning,ieracitano2021novel,bang2021spatio,masse2022classification,hammer2022interpretable,ravindran2022decoding,du2022ienet,banville2022robust,tajmirriahi2022interpretable,van2022classification,niu2022knowledge,bang2022interpretable,stuart2022interpretable,salami2022eeg,mcdermott2021artifacts,choi2021non,dong2021explainable,petrosyan2021compact,kobler2021interpretation,lekova2021fuzzy,fu2021recognizing,galindo2020multiple,jin2020interpretable,borra2020interpretable,xu2020tangent,huang2020spectrum,loza2019discrimination,lopez2019supervised,rahimi2018efficient,motrenko2018multi,williams2018unsupervised,bouchard2017sparse,caywood2017gaussian,sturm2016interpretable,wang2016unsupervised,dyson2015online,mametkulov2022explainable,dong2023heterogeneous,cui2022compact,zhao2023signal,kim2021deep,cui2022eeg,kia2017brain,lee2022quantifying,kuang2023seer,thanigaivelu2023oisvm,tan2023eeg,tanaka2020group,hu2020assessment,kostas2019machine,collazos2020cnn,vidaurre2023identification,park2022individualized,jiang2020smart,raab2022xai4eeg,petrescu2021machine,svetlakov2022representation,lo2022permutation,na2022objective,zhang2020eeg,hsieh2021explainable,gabeff2021interpreting}\end{tabular} \\ \hline 
  
% \begin{tabular}[c]{@{}l@{}}\rotatebox[origin=c]{90}{\parbox{2.5cm}{Knowledge \\ Discovery}}\end{tabular} & \begin{tabular}[c]{{ @{} p{\dimexpr0.95\textwidth} @{} }}
\begin{tabular}[c]{@{}l@{}}Knowledge \\ Discovery\end{tabular} & \begin{tabular}[c]{{ @{} p{\dimexpr0.95\textwidth} @{} }}
\cite{urdanetaexplainable,kim2023designing,nagarajan2022relevance,kim2023identification,petrosyan2021decoding,fu2022single,borra2022bayesian,ieracitano2021novel,hammer2022interpretable,ravindran2022decoding,apicella2022toward,choi2021non,dong2021explainable,aellen2021convolutional,fu2021recognizing,borra2020interpretable,motrenko2018multi,williams2018unsupervised,bastos2016discovering,dyson2015online,dong2023heterogeneous,cui2022compact,cui2022eeg,lee2022quantifying,kuang2023seer,tanaka2020group,hu2020assessment,vidaurre2023identification,park2022individualized}\end{tabular} \\ \hline

% \begin{tabular}[c]{@{}l@{}}\rotatebox[origin=c]{90}{\parbox{5cm}{Model Improvement \\ or Debugging}}\end{tabular} & \begin{tabular}[c]{{ @{} p{\dimexpr0.95\textwidth} @{} }}
\begin{tabular}[c]{@{}l@{}}Model \\ Improvement \\ or Debugging\end{tabular} & \begin{tabular}[c]{{ @{} p{\dimexpr0.95\textwidth} @{} }}
\cite{urdanetaexplainable,song2022eeg,dutt2022sleepxai,xie2022transformer,giudice2022visual,kim2023designing,nagarajan2022relevance,kim2023identification,shibu2022explainable,petrosyan2022speech,petrosyan2021decoding,kumar2022neurophysiologically,fu2022single,chen2022novel,lawhern2018eegnet,zygierewicz2022decoding,islam2022explainable,borra2022bayesian,zhao2019learning,ieracitano2021novel,bang2021spatio,masse2022classification,hammer2022interpretable,ravindran2022decoding,du2022ienet,banville2022robust,tajmirriahi2022interpretable,van2022classification,niu2022knowledge,bang2022interpretable,apicella2022toward,stuart2022interpretable,qu2022eeg4home,salami2022eeg,mcdermott2021artifacts,choi2021non,dong2021explainable,aellen2021convolutional,petrosyan2021compact,kobler2021interpretation,lekova2021fuzzy,fu2021recognizing,galindo2020multiple,jin2020interpretable,borra2020interpretable,xu2020tangent,huang2020spectrum,loza2019discrimination,lopez2019supervised,rahimi2018efficient,motrenko2018multi,williams2018unsupervised,bouchard2017sparse,caywood2017gaussian,sturm2016interpretable,wang2016unsupervised,dyson2015online,mametkulov2022explainable,kim2021deep,kia2017brain,kuang2023seer,thanigaivelu2023oisvm,tanaka2020group,collazos2020cnn}\end{tabular} \\ \hline

\begin{tabular}[c]{@{}l@{}}Usability \\ and Control\end{tabular} & \begin{tabular}[c]{{ @{} p{\dimexpr0.95\textwidth} @{} }}
\cite{qu2022eeg4home,lekova2021fuzzy}\end{tabular} \\ \hline
\end{longtable}
% \end{landscape}
% \twocolumn

\setlength\LTleft{0cm}\begin{longtable}{ll p{12cm}}
\caption{XAI4BCI Design space variable: "What?", its categorisation and corresponding references from current literature}
\label{tab:DSWhat}\\
\hline
\multicolumn{1}{c}{\textbf{Categories}} & \multicolumn{1}{c}{\textbf{References}} \\ \hline
\endhead
\begin{tabular}[c]{@{}l@{}}Motor \\ Movements \\ or \\ Imagery\end{tabular} & \begin{tabular}[c]{{ @{} p{\dimexpr0.95\textwidth} @{} }}
% \multicolumn{1}{c}{\multirow{8}{*}[-12em]{What?}} & \begin{tabular}[c]{@{}l@{}}Motor \\ Movements \\ or \\ Imagery\end{tabular} & \begin{tabular}[c]{{ @{} p{\dimexpr0.85\textwidth} @{} }}
\cite{song2022eeg,xie2022transformer,kim2023designing,nagarajan2022relevance,kim2023identification,shibu2022explainable,petrosyan2021decoding,kumar2022neurophysiologically,fu2022single,lawhern2018eegnet,zhao2019learning,ieracitano2021novel,bang2021spatio,hammer2022interpretable,ravindran2022decoding,du2022ienet,niu2022knowledge,bang2022interpretable,qu2022eeg4home,salami2022eeg,mcdermott2021artifacts,choi2021non,dong2021explainable,kobler2021interpretation,fu2021recognizing,galindo2020multiple,borra2020interpretable,xu2020tangent,huang2020spectrum,loza2019discrimination,lopez2019supervised,rahimi2018efficient,motrenko2018multi,williams2018unsupervised,sturm2016interpretable,wang2016unsupervised,dyson2015online,dong2023heterogeneous,kim2021deep,thanigaivelu2023oisvm,collazos2020cnn,svetlakov2022representation,hsieh2021explainable}\end{tabular} \\ \hline 

\begin{tabular}[c]{@{}l@{}}Visual \\ Stimuli\end{tabular} & \begin{tabular}[c]{{ @{} p{\dimexpr0.95\textwidth} @{} }}
\cite{chen2022novel,lawhern2018eegnet,islam2022explainable,borra2022bayesian,du2022ienet,tajmirriahi2022interpretable,van2022classification,qu2022eeg4home,aellen2021convolutional,lekova2021fuzzy,caywood2017gaussian,kim2021deep,kia2017brain,tan2023eeg,tanaka2020group}\end{tabular} \\ \hline 

\begin{tabular}[c]{@{}l@{}}Attention \\ and \\ Memory\end{tabular} & \begin{tabular}[c]{{ @{} p{\dimexpr0.95\textwidth} @{} }}
\cite{zygierewicz2022decoding,du2022ienet,qu2022eeg4home,kobler2021interpretation,lekova2021fuzzy,caywood2017gaussian,mametkulov2022explainable,cui2022compact,cui2022eeg}\end{tabular} \\ \hline

\begin{tabular}[c]{@{}l@{}}Audio \\ Decoding\end{tabular} & \begin{tabular}[c]{{ @{} p{\dimexpr0.95\textwidth} @{} }}
\cite{masse2022classification,aellen2021convolutional,lekova2021fuzzy,caywood2017gaussian,kim2021deep,na2022objective,zhang2020eeg}\end{tabular} \\ \hline 

\begin{tabular}[c]{@{}l@{}}Speech \\ Decoding\end{tabular} & \begin{tabular}[c]{{ @{} p{\dimexpr0.95\textwidth} @{} }}
\cite{petrosyan2022speech,stuart2022interpretable,petrosyan2021compact,bouchard2017sparse,wang2016unsupervised,kostas2019machine}\end{tabular} \\ \hline 

\begin{tabular}[c]{@{}l@{}}Emotion \\ Classification\end{tabular} & \begin{tabular}[c]{{ @{} p{\dimexpr0.95\textwidth} @{} }}
\cite{song2022eeg,apicella2022toward,jin2020interpretable,kuang2023seer,petrescu2021machine}\end{tabular} \\ \hline 

\begin{tabular}[c]{@{}l@{}}Sleep Stage \\ Classification\end{tabular} & \begin{tabular}[c]{{ @{} p{\dimexpr0.95\textwidth} @{} }}
\cite{dutt2022sleepxai,banville2022robust,zhao2023signal,hu2020assessment}\end{tabular} \\ \hline 

Others & \begin{tabular}[c]{{ @{} p{\dimexpr0.95\textwidth} @{} }}
\cite{urdanetaexplainable,giudice2022visual,bastos2016discovering,lee2022quantifying,hu2020assessment,vidaurre2023identification,park2022individualized,jiang2020smart,lo2022permutation,gabeff2021interpreting}\end{tabular} \\ \hline
\end{longtable}
% \end{landscape}
% \twocolumn

\setlength\LTleft{0cm}\begin{longtable}{ll p{12cm}}
\caption{XAI4BCI Design space variable: \textit{How? (XAI Methods)}, its categorisation and corresponding references from current literature}
\label{tab:DSHowMethods}\\
\hline
\multicolumn{1}{c}{\textbf{Categories}} & \multicolumn{1}{c}{\textbf{References}} \\ \hline
\endhead
% \noalign{\penalty-10000}
\begin{tabular}[c]{@{}l@{}}SHAP, \\ DeepSHAP, \\ DeepLIFT\end{tabular} & \begin{tabular}[c]{{ @{} p{\dimexpr0.95\textwidth} @{} }}
\cite{urdanetaexplainable,kim2023designing,shibu2022explainable,islam2022explainable,du2022ienet,apicella2022toward,raab2022xai4eeg,gabeff2021interpreting}\end{tabular} \\ \hline 
 \begin{tabular}[c]{@{}l@{}}Integrated \\ Gradients, \\ LRP\end{tabular} & \begin{tabular}[c]{{ @{} p{\dimexpr0.95\textwidth} @{} }}
\cite{nagarajan2022relevance,kim2023identification,lawhern2018eegnet,bang2021spatio,du2022ienet,bang2022interpretable,apicella2022toward,dong2021explainable,sturm2016interpretable,lee2022quantifying,svetlakov2022representation,zhang2020eeg}\end{tabular} \\ \hline 
 \begin{tabular}[c]{@{}l@{}}Saliency \\ maps, CAM, \\ Grad CAM\end{tabular} & \begin{tabular}[c]{{ @{} p{\dimexpr0.95\textwidth} @{} }}
\cite{song2022eeg,dutt2022sleepxai,giudice2022visual,kumar2022neurophysiologically,lawhern2018eegnet,ravindran2022decoding,du2022ienet,tajmirriahi2022interpretable,apicella2022toward,choi2021non,aellen2021convolutional,cui2022compact,zhao2023signal,cui2022eeg,kuang2023seer,tan2023eeg,kostas2019machine}\end{tabular} \\ \hline 
\begin{tabular}[c]{@{}l@{}}Attention, \\ Frequency \\ Filter \\ Weight \\ Visualisation\end{tabular} & \begin{tabular}[c]{{ @{} p{\dimexpr0.95\textwidth} @{} }}
\cite{xie2022transformer,petrosyan2022speech,petrosyan2021decoding,chen2022novel,lawhern2018eegnet,borra2022bayesian,zhao2019learning,du2022ienet,banville2022robust,van2022classification,niu2022knowledge,apicella2022toward,stuart2022interpretable,salami2022eeg,mcdermott2021artifacts,borra2020interpretable,xu2020tangent,huang2020spectrum,bouchard2017sparse,kia2017brain,kuang2023seer,collazos2020cnn,park2022individualized,hsieh2021explainable}\end{tabular} \\ \hline 
\begin{tabular}[c]{@{}l@{}}Others \\ (LIME, Tree, \\ Tsne,\\ Occlusion \\ Sensitivity)\end{tabular} & \begin{tabular}[c]{{ @{} p{\dimexpr0.95\textwidth} @{} }}
\cite{urdanetaexplainable,song2022eeg,xie2022transformer,giudice2022visual,fu2022single,zygierewicz2022decoding,ieracitano2021novel,masse2022classification,hammer2022interpretable,du2022ienet,qu2022eeg4home,petrosyan2021compact,kobler2021interpretation,lekova2021fuzzy,fu2021recognizing,galindo2020multiple,jin2020interpretable,loza2019discrimination,lopez2019supervised,rahimi2018efficient,motrenko2018multi,williams2018unsupervised,caywood2017gaussian,wang2016unsupervised,bastos2016discovering,dyson2015online,mametkulov2022explainable,dong2023heterogeneous,kim2021deep,thanigaivelu2023oisvm,tanaka2020group,hu2020assessment,vidaurre2023identification,jiang2020smart,petrescu2021machine,lo2022permutation,na2022objective}\end{tabular} \\ \hline
\end{longtable}
% \end{landscape}
% \twocolumn

\setlength\LTleft{0cm}\begin{longtable}{ll p{12cm}}
\caption{XAI4BCI Design space variable: "How? (Interfaces)", its categorisation and corresponding references from current literature}
\label{tab:DSHowInterfaces}\\
\hline
\multicolumn{1}{c}{\textbf{Categories}} & \multicolumn{1}{c}{\textbf{References}} \\ \hline
\endhead
% \noalign{\penalty-10000}
\begin{tabular}[c]{@{}l@{}}Spatial \\ Patterns \\ (Topographic \\ Maps)\end{tabular} & \begin{tabular}[c]{{ @{} p{\dimexpr0.95\textwidth} @{} }}
\cite{song2022eeg,xie2022transformer,kim2023designing,nagarajan2022relevance,kim2023identification,shibu2022explainable,petrosyan2022speech,petrosyan2021decoding,kumar2022neurophysiologically,chen2022novel,lawhern2018eegnet,borra2022bayesian,zhao2019learning,ieracitano2021novel,bang2021spatio,hammer2022interpretable,ravindran2022decoding,du2022ienet,banville2022robust,van2022classification,bang2022interpretable,salami2022eeg,mcdermott2021artifacts,choi2021non,aellen2021convolutional,kobler2021interpretation,galindo2020multiple,borra2020interpretable,xu2020tangent,huang2020spectrum,loza2019discrimination,bouchard2017sparse,caywood2017gaussian,sturm2016interpretable,wang2016unsupervised,dyson2015online,mametkulov2022explainable,kim2021deep,cui2022eeg,kia2017brain,kuang2023seer,tan2023eeg,tanaka2020group,hu2020assessment,kostas2019machine,collazos2020cnn,vidaurre2023identification,park2022individualized,raab2022xai4eeg,lo2022permutation,na2022objective,zhang2020eeg,hsieh2021explainable}\end{tabular} \\ \hline 
\begin{tabular}[c]{@{}l@{}}Temporal \\ Patterns\end{tabular} & \begin{tabular}[c]{{ @{} p{\dimexpr0.95\textwidth} @{} }}
\cite{dutt2022sleepxai,giudice2022visual,kim2023designing,kumar2022neurophysiologically,chen2022novel,lawhern2018eegnet,ieracitano2021novel,ravindran2022decoding,du2022ienet,tajmirriahi2022interpretable,van2022classification,niu2022knowledge,qu2022eeg4home,mcdermott2021artifacts,aellen2021convolutional,loza2019discrimination,williams2018unsupervised,bouchard2017sparse,dyson2015online,cui2022compact,zhao2023signal,cui2022eeg,kia2017brain,kuang2023seer,tan2023eeg,tanaka2020group,kostas2019machine,raab2022xai4eeg,svetlakov2022representation,hsieh2021explainable,gabeff2021interpreting}\end{tabular} \\ \hline 
\begin{tabular}[c]{@{}l@{}}Feature \\ Relevance \\ Map\end{tabular} & \begin{tabular}[c]{{ @{} p{\dimexpr0.95\textwidth} @{} }}
\cite{urdanetaexplainable,kim2023designing,shibu2022explainable,fu2022single,lawhern2018eegnet,zygierewicz2022decoding,islam2022explainable,ravindran2022decoding,du2022ienet,banville2022robust,van2022classification,apicella2022toward,dong2021explainable,petrosyan2021compact,kobler2021interpretation,fu2021recognizing,galindo2020multiple,jin2020interpretable,lopez2019supervised,rahimi2018efficient,motrenko2018multi,williams2018unsupervised,bouchard2017sparse,dong2023heterogeneous,cui2022compact,zhao2023signal,cui2022eeg,lee2022quantifying,kuang2023seer,petrescu2021machine,svetlakov2022representation,na2022objective,zhang2020eeg,hsieh2021explainable,gabeff2021interpreting}\end{tabular} \\ \hline 
\begin{tabular}[c]{@{}l@{}}Spectral \\ Profiles\end{tabular} & \begin{tabular}[c]{{ @{} p{\dimexpr0.95\textwidth} @{} }}
\cite{petrosyan2022speech,petrosyan2021decoding,lawhern2018eegnet,borra2022bayesian,zhao2019learning,bang2021spatio,hammer2022interpretable,du2022ienet,niu2022knowledge,stuart2022interpretable,salami2022eeg,mcdermott2021artifacts,petrosyan2021compact,galindo2020multiple,borra2020interpretable,huang2020spectrum,loza2019discrimination,lopez2019supervised,motrenko2018multi,dyson2015online,cui2022eeg,thanigaivelu2023oisvm,tanaka2020group,hu2020assessment,kostas2019machine,jiang2020smart,raab2022xai4eeg,svetlakov2022representation}\end{tabular} \\ \hline
\end{longtable}
% \end{landscape}
% \twocolumn

\section{Discussion}
\label{sec:discussion}

Our work complements the design space proposed by \citet{kosmyna2019conceptual}. However, their discussion focuses on electroencephalography (EEG) based BCI and the levels of abstraction that form the basis of the "How?" axis. \citet{mason2003general} discusses the taxonomy and general framework dating back to the inception of BCI. Our work augments the previous works, illustrating the growth of evolving design spaces. 

BCIs require humans in the loop: humans as users and humans who check and analyse BCI performance. The communication between the brain and machines requires efficient communication, trust, clarity, and understanding. XAI4BCI can address such challenges by combining symbolic AI and Machine Learning. For example, hybrid models can incorporate both numerical and symbolic components. A neural network can be used for feature extraction and numerical processing and then connected to a symbolic reasoning engine that interprets the results and makes decisions based on logical rules. Different communities of AI have studied such topics with different definitions, evaluation metrics, motivations, and results. Therefore, we discuss various perspectives to approach XAI4BCI.

\subsection{XAI as an engineering tool for BCI}
Understanding the duality of BCI applications that seek high efficiency for their applications vis-a-vis understanding of neurophysiological mechanisms, XAI4BCI serves a common thread. Considering the engineering applications of BCI for assistive and recreational technologies, there are motivations to seek explanations to optimise the control in the environment and improve the model performance. While, on the scientific front, BCI for diagnosis and rehabilitation seeks explanations to justify the model predictions and discover the neurophysiological mechanisms. This intersection highlights the essential overlap of domain knowledge and engineering. \citet{zhang2021tiny,meng2023adversarial,zhou2023interpretable} report the impact of adversarial noise on models used for BCI. XAI has played a crucial role in investigating robustness from adversarial noise in the models for other domains \citep{etmann2019connection,ross2018improving,Chan_2020_CVPR}. \citet{weber2022beyond} propose approaches to leverage explanations for improving models by augmenting different components of the training process or by adapting the trained model. The study also includes empirical experiments showing how explanations can help model generalization ability or reasoning. 
However, the accuracy-explainability trade-off often prioritises accuracy in engineering applications. 

\subsection{XAI as a research tool for BCI}
Whereas BCIs for diagnosis and rehabilitation need the interpretations to justify the model predictions or discover the neural mechanisms relevant for the dynamics of brain responses, \citet{kim2023identification} use XAI to identify the cerebral cortices processing kinematic parameters (i.e., acceleration, velocity, and position) during reaching movement from magnetoencephalography (MEG) data. XAI facilitates an additional medium apart from invasive single neuron studies or large-scale coherence studies to investigate the cortical areas involved in decoding the kinematic parameters. The diverse approaches of BCI for engineering and science motivate the inclusion of XAI by considering the context and citing the significance of a consolidated design space.

\subsection{XAI beyond model interpretation}

Models or systems whose decisions cannot be well interpreted can be hard to accept~\citep{reddy2022explainability}. An explanation of the BCI performance is needed for good and bad performance. If an individual cannot control a BCI, it is necessary to explain to the user (and family) why the BCI does not work; there should not be this notion of chance where BCI may or may not work for a given individual. Likewise, it is necessary to explain why the performance is high and why it stops working when it should, including the hardware, such as sensor disconnections or malfunctions.

\begin{figure}[t!]
\begin{center}
\includegraphics[width=\linewidth]{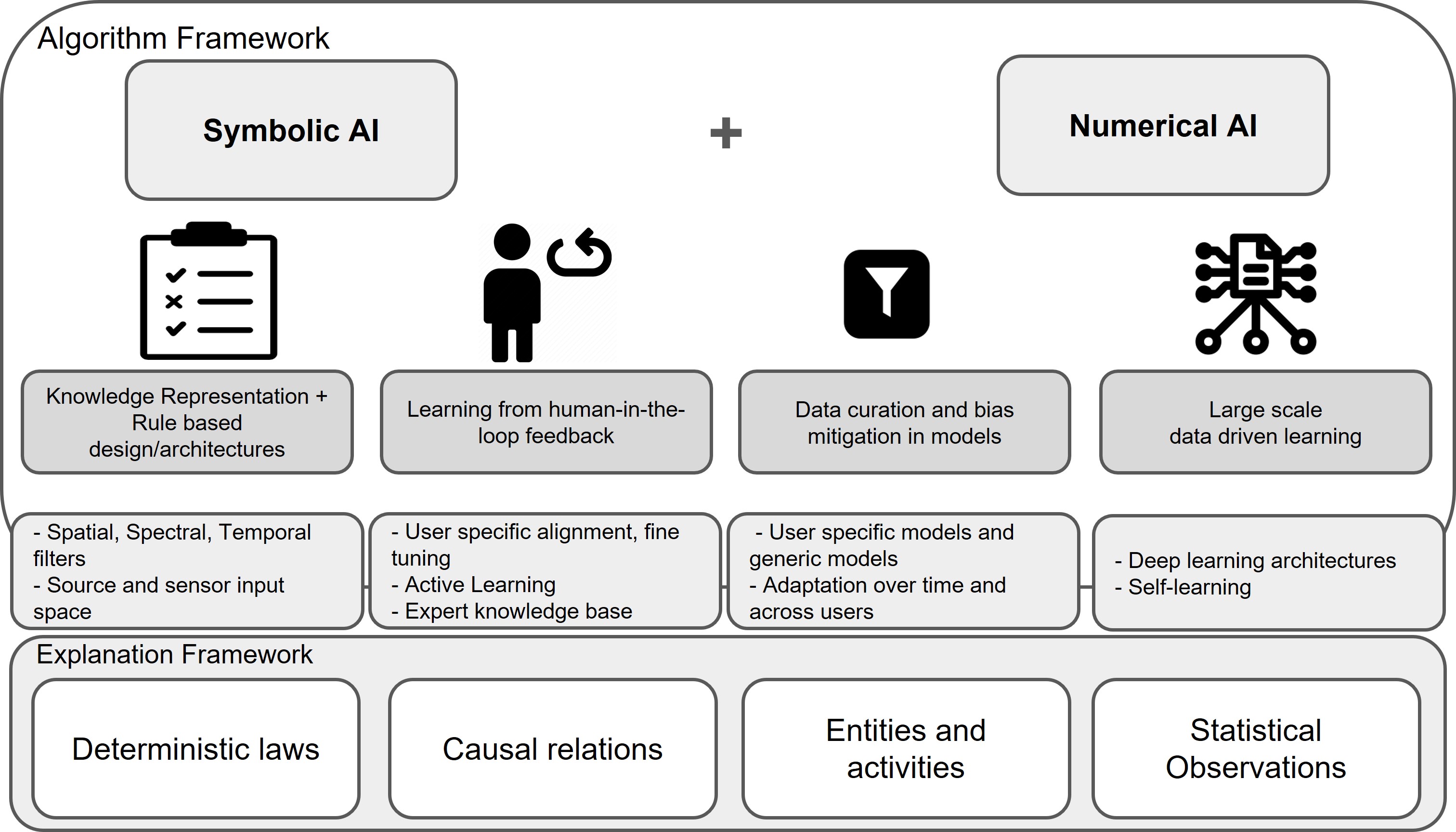}
\caption{Algorithm and explanation framework comprehensively aides XAI4BCI.}
\label{fig:symAInumAI}
\end{center}
\end{figure}

While our design space XAI4BCI proposes the inclusion of deterministic laws, causal relations, and mechanisms in the explanation framework, there are seldom any discussions to formulate a domain-specific knowledge base into XAI. This research direction remains to be explored. Inculcating causal relations by defining or learning specific spatial, spectral, and temporal filters contributes to the explanation framework. Similarly, adding knowledge representations, curating data via feedback from human-in-the-loop, and learning to mitigate bias in models and data aid a strong explanation framework. \autoref{fig:symAInumAI} comprehensively represents the underlying frameworks proposed in XAI4BCI and the techniques that could assist in realising the explanation framework via evolution in the algorithm framework. 

Once the explanation is obtained, it is necessary to visualize the explanations in a clear manner. In particular, when using graph representation, stakeholders must access the explanation of the decisions in a clear way that takes into account the best practices in user experience and data visualization techniques. 
% To work on this paragraph, --> idea: explanation is not the end goal, it is about how to communicate the explanation to the stakeholder so they can make informed decisions

\subsection{Pitfalls of XAI4BCI}
Current XAI techniques also have shortcomings, \citet{ehsan2021explainability} categorize them into two based on the intentionality: dark patterns \citep{anders2020fairwashing,chromik2019dark} and pitfalls. Dark patterns are intended to harm users by manipulating explanations, while pitfalls are unrealized intellectual blind spots leading to side-effects from XAI. Their work also proposes strategies at the research, design, and organisational levels to overcome pitfalls. Studies on the conceptual and technical front are necessary for evaluation, awareness among stakeholders and to overcome possible shortcomings in XAI~\citep{kumar2020problems, verdinelli2023feature,dombrowski2019explanations,han2022explanation}. 

\citet{haufe2014interpretation} discusses a common misconception in BCI research regarding the interpretability of decoding models. They highlight that while decoding models are designed to estimate brain states to control BCIs, there is often an implicit assumption that the parameters of these models can be easily interpreted in terms of the properties of the brain. They emphasize the potential risks associated with misinterpreting the weights of multivariate classifiers, especially when making decisions based on those interpretations. The importance of raising awareness within the neuroimaging community about the potential pitfalls of misinterpreting classifier weights has been discussed even for simpler linear models. In agreement with their motivation, improving the interpretability of BCI models is essential, and the stakeholders should strive for meaningful and accurate explanations to avoid potential errors and consequences.

\subsection{Limitations of the study}

Recognizing that the increasing demand for explainability and interpretability emerged with the rise of end-to-end prediction models, we considered that studies from the post-2015 period significantly encompass the progress in this area. This consideration is influenced by \citet{sturm2016interpretable} in 2016 and the development of the first successful large-scale model like EEGNet \citep{lawhern2018eegnet} in 2018. We acknowledge that handcrafted feature extraction pipelines are transparent and trustworthy, suitable for comparisons and benchmarks. Our work may exclude relevant interpretable or explainable approaches published before 2015. For addressing RQ1 and RQ2, the categorization in the study was subjective, relying on specific keywords to identify the motivation and purpose for the study.

Discussing the broader limitations of the current framework for explanations, the common human tendency to over-trust a given explanation derived from a limited dataset can be a definitive risk \citep{howard2020we}. Hence, it is essential to guard the stakeholders against the risk of introducing biases into decision-making \citep{ghassemi2021false}. By acknowledging this fact, the design space needs to evolve, and explanation interfaces must highlight this cautionary note. For example, acknowledging that spatial maps indicate the locations where the model is weighed upon for the prediction does not resolve the gap of whether it was appropriate to weigh upon those brain regions for the prediction. This is left to the stakeholders to identify and may lead to mistakes.

XAI4BCI assists in developing visualizations and extracting explanations for the predictions. XAI4BCI helps standardize the comparisons of explanations from various models using different techniques and interfaces; however, how to evaluate them objectively and developing relevant metrics is an essential direction for future research.

\section{Conclusion}
\label{sec:conclusion}

The primary goal of Brain-Computer Interface (BCI) systems should extend beyond mere performance metrics. While performance is crucial, it is just one aspect contributing to the usability of the system beyond the controlled environment of a laboratory. A key requirement is the ability to explain and comprehend the system's performance, whether functioning as intended or encountering issues. Attaining explainability in BCIs is a formidable yet essential task to instill trust in predictive models and assist stakeholders in making informed decisions.

In BCI, the fusion of neuroscience knowledge directly into pattern recognition application architectures provides guidance for processing and analyzing signals. This encompasses decisions about sensor selection, frequency bands, and the viability of spatial filters. Concurrently, deep learning classifiers, which process raw EEG signals, necessitate analysis to elucidate the rationale behind different layers and learned spatial/temporal filters. The pursuit of explainability should not be confined to the software domain alone; understanding and explaining issues at the hardware level is critical once a BCI is deployed in a clinical environment or at home.

The emerging trend of Explainable AI for Brain-Computer Interfaces (XAI4BCI) aligns with the contemporary movement in AI to integrate both symbolic and numerical AI approaches. While numerical AI has historically been the predominant focus, the inclusion of symbolic AI is essential for constructing more resilient systems capable of symbiotic interaction with humans. The meta-analysis of the current BCI literature and the proposed design space for XAI4BCI considers neurophysiological foundations and encourages dialogue on establishing standards for incorporating explanations in BCIs. This multifaceted approach aims to enhance performance and communication with diverse audiences, laying the groundwork for advancements in the field of XAI4BCI.

\section{Acknowledgment}
\label{sec:acknowledgment}
All source code data is available on \href{https://github.com/MIILab-IITGN/XAI4BCI}{https://github.com/MIILab-IITGN/XAI4BCI}. This work was supported by Indian Institute of Technology Gandhinagar startup grant IP/IITGN/CSE/YM/2324/05.  

\section{Author contributions}
\label{sec:author_contributions}
PR and YKM contributed to conceptualization work, data curation, formal analysis, and visualization. PR and YKM wrote the first draft of the manuscript. HC and YKM validated the work. All authors reviewed, edited, and approved the final manuscript.

%%%%%%%%%%%%%%%%%%%%%%%%%%%%%%%%%%%%%
\newcommand{\newblock}{}
\bibliographystyle{dcu}
\bibliography{references}

@article{page2021prisma,
  title={PRISMA 2020 explanation and elaboration: updated guidance and exemplars for reporting systematic reviews},
  author={Page, Matthew J and Moher, David and Bossuyt, Patrick M and Boutron, Isabelle and Hoffmann, Tammy C and Mulrow, Cynthia D and Shamseer, Larissa and Tetzlaff, Jennifer M and Akl, Elie A and Brennan, Sue E and others},
  journal={bmj},
  volume={372},
  year={2021},
  publisher={British Medical Journal Publishing Group}
}

@article{prisma2021,
  title={The PRISMA 2020 statement: an updated guideline for reporting systematic reviews},
  author={Page, Matthew J and McKenzie, Joanne E and Bossuyt, Patrick M and Boutron, Isabelle and Hoffmann, Tammy C and Mulrow, Cynthia D and Shamseer, Larissa and Tetzlaff, Jennifer M and Akl, Elie A and Brennan, Sue E and others},
  journal={International journal of surgery},
  volume={88},
  pages={105906},
  year={2021},
  publisher={Elsevier}
}

@article{london2019artificial,
  title={Artificial intelligence and black-box medical decisions: accuracy versus explainability},
  author={London, Alex John},
  journal={Hastings Center Report},
  volume={49},
  number={1},
  pages={15--21},
  year={2019},
  publisher={Wiley Online Library}
}

@article{arrieta2020explainable,
  title={Explainable Artificial Intelligence (XAI): Concepts, taxonomies, opportunities and challenges toward responsible AI},
  author={Arrieta, Alejandro Barredo and D{\'\i}az-Rodr{\'\i}guez, Natalia and Del Ser, Javier and Bennetot, Adrien and Tabik, Siham and Barbado, Alberto and Garc{\'\i}a, Salvador and Gil-L{\'o}pez, Sergio and Molina, Daniel and Benjamins, Richard and others},
  journal={Information fusion},
  volume={58},
  pages={82--115},
  year={2020},
  publisher={Elsevier}
}

@book{walter2008cambridge,
  title={Cambridge Advanced Learner's Dictionary with CD-ROM},
  author={Walter, Elizabeth},
  year={2008},
  publisher={Cambridge university press}
}

@article{erasmus2021interpretability,
  title={What is interpretability?},
  author={Erasmus, Adrian and Brunet, Tyler DP and Fisher, Eyal},
  journal={Philosophy \& Technology},
  volume={34},
  number={4},
  pages={833--862},
  year={2021},
  publisher={Springer}
}

@article{erasmus2022interpretability,
  title={Interpretability and unification},
  author={Erasmus, Adrian and Brunet, Tyler DP},
  journal={Philosophy \& Technology},
  volume={35},
  number={2},
  pages={42},
  year={2022},
  publisher={Springer}
}

@article{ding2022explainability,
  title={Explainability of artificial intelligence methods, applications and challenges: A comprehensive survey},
  author={Ding, Weiping and Abdel-Basset, Mohamed and Hawash, Hossam and Ali, Ahmed M},
  journal={Information Sciences},
  year={2022},
  publisher={Elsevier}
}

@article{miller2019explanation,
  title={Explanation in artificial intelligence: Insights from the social sciences},
  author={Miller, Tim},
  journal={Artificial intelligence},
  volume={267},
  pages={1--38},
  year={2019},
  publisher={Elsevier}
}

@article{siddiqi2022causal,
  title={Causal mapping of human brain function},
  author={Siddiqi, Shan H and Kording, Konrad P and Parvizi, Josef and Fox, Michael D},
  journal={Nature reviews neuroscience},
  volume={23},
  number={6},
  pages={361--375},
  year={2022},
  publisher={Nature Publishing Group UK London}
}

@article{castelvecchi2016can,
  title={Can we open the black box of AI?},
  author={Castelvecchi, Davide},
  journal={Nature News},
  volume={538},
  number={7623},
  pages={20},
  year={2016}
}

@article{tjoa2020survey,
  title={A survey on explainable artificial intelligence (xai): Toward medical xai},
  author={Tjoa, Erico and Guan, Cuntai},
  journal={IEEE transactions on neural networks and learning systems},
  volume={32},
  number={11},
  pages={4793--4813},
  year={2020},
  publisher={IEEE}
}

@article{doshi2017towards,
  title={Towards a rigorous science of interpretable machine learning},
  author={Doshi-Velez, Finale and Kim, Been},
  journal={arXiv preprint arXiv:1702.08608},
  year={2017}
}

@inproceedings{tonekaboni2019clinicians,
  title={What clinicians want: contextualizing explainable machine learning for clinical end use},
  author={Tonekaboni, Sana and Joshi, Shalmali and McCradden, Melissa D and Goldenberg, Anna},
  booktitle={Machine learning for healthcare conference},
  pages={359--380},
  year={2019},
  organization={PMLR}
}

@inproceedings{ribeiro2016should,
  title={" Why should i trust you?" Explaining the predictions of any classifier},
  author={Ribeiro, Marco Tulio and Singh, Sameer and Guestrin, Carlos},
  booktitle={Proceedings of the 22nd ACM SIGKDD international conference on knowledge discovery and data mining},
  pages={1135--1144},
  year={2016}
}

@book{de2009scientific,
  title={Scientific understanding: Philosophical perspectives},
  author={De Regt, Henk W and Leonelli, Sabina and Eigner, Kai},
  year={2009},
  publisher={University of Pittsburgh Pre}
}

@book{strevens2011depth,
  title={Depth: An account of scientific explanation},
  author={Strevens, Michael},
  year={2011},
  publisher={Harvard University Press}
}

@article{potochnik2016scientific,
  title={Scientific explanation: Putting communication first},
  author={Potochnik, Angela},
  journal={Philosophy of Science},
  volume={83},
  number={5},
  pages={721--732},
  year={2016},
  publisher={Cambridge University Press}
}

@article{strevens2013no,
  title={No understanding without explanation},
  author={Strevens, Michael},
  journal={Studies in history and philosophy of science Part A},
  volume={44},
  number={3},
  pages={510--515},
  year={2013},
  publisher={Elsevier}
}

@book{de2017understanding,
  title={Understanding scientific understanding},
  author={De Regt, Henk W},
  year={2017},
  publisher={Oxford University Press}
}

@book{khalifa2017understanding,
  title={Understanding, explanation, and scientific knowledge},
  author={Khalifa, Kareem},
  year={2017},
  publisher={Cambridge University Press}
}

@article{gunning2017explainable,
  title={Explainable artificial intelligence (xai)},
  author={Gunning, David},
  journal={Defense advanced research projects agency (DARPA), nd Web},
  volume={2},
  number={2},
  pages={1},
  year={2017}
}

@article{urdanetaexplainable,
  title={Explainable Machine Learning Predictions for the Long-term Performance of Brain-Computer Interfaces},
  year={2023},
  author={Urdaneta, Morgan E and Veit, Nicole C and Burke, Renae G and Malone, Ian G and Smith, Kaleb E and Otto, Kevin J}
}

@article{song2022eeg,
  title={EEG Conformer: Convolutional Transformer for EEG Decoding and Visualization},
  author={Song, Yonghao and Zheng, Qingqing and Liu, Bingchuan and Gao, Xiaorong},
  journal={IEEE Transactions on Neural Systems and Rehabilitation Engineering},
  year={2022},
  publisher={IEEE}
}

@article{dutt2022sleepxai,
  title={SleepXAI: An explainable deep learning approach for multi-class sleep stage identification},
  author={Dutt, Micheal and Redhu, Surender and Goodwin, Morten and Omlin, Christian W},
  journal={Applied Intelligence},
  pages={1--14},
  year={2022},
  publisher={Springer}
}

@article{xie2022transformer,
  title={A Transformer-based approach combining deep learning network and spatial-temporal information for raw EEG classification},
  author={Xie, Jin and Zhang, Jie and Sun, Jiayao and Ma, Zheng and Qin, Liuni and Li, Guanglin and Zhou, Huihui and Zhan, Yang},
  journal={IEEE Transactions on Neural Systems and Rehabilitation Engineering},
  volume={30},
  pages={2126--2136},
  year={2022},
  publisher={IEEE}
}

@inproceedings{giudice2022visual,
  title={Visual Explanations of Deep Convolutional Neural Network for eye blinks detection in EEG-based BCI applications},
  author={Giudice, Michele Lo and Mammone, Nadia and Ieracitano, Cosimo and Campolo, Maurizio and Bruna, Arcangelo Ranieri and Tomaselli, Valeria and Morabito, Francesco Carlo},
  booktitle={2022 International Joint Conference on Neural Networks (IJCNN)},
  pages={01--08},
  year={2022},
  organization={IEEE}
}

@article{kim2023designing,
  title={Designing an XAI interface for BCI experts: A contextual design for pragmatic explanation interface based on domain knowledge in a specific context},
  author={Kim, Sangyeon and Choo, Sanghyun and Park, Donghyun and Park, Hoonseok and Nam, Chang S and Jung, Jae-Yoon and Lee, Sangwon},
  journal={International Journal of Human-Computer Studies},
  volume={174},
  pages={103009},
  year={2023},
  publisher={Elsevier}
}

@article{nagarajan2022relevance,
  title={Relevance based channel selection in motor imagery brain-computer interface},
  author={Nagarajan, Aarthy and Robinson, Neethu and Guan, Cuntai},
  journal={Journal of Neural Engineering},
  year={2022}
}

@article{kim2023identification,
  title={Identification of cerebral cortices processing acceleration, velocity, and position during directional reaching movement with deep neural network and explainable AI},
  author={Kim, HongJune and Kim, June Sic and Chung, Chun Kee},
  journal={NeuroImage},
  volume={266},
  pages={119783},
  year={2023},
  publisher={Elsevier}
}

@article{shibu2022explainable,
  title={Explainable artificial intelligence model to predict brain states from fNIRS signals},
  author={Shibu, Caleb Jones and Sreedharan, Sujesh and Arun, KM and Kesavadas, Chandrasekharan and Sitaram, Ranganatha},
  journal={Frontiers in Human Neuroscience},
  volume={16},
  year={2022},
  publisher={Frontiers Media SA}
}

@article{petrosyan2022speech,
  title={Speech decoding from a small set of spatially segregated minimally invasive intracranial EEG electrodes with a compact and interpretable neural network},
  author={Petrosyan, Artur and Voskoboinikov, Alexey and Sukhinin, Dmitrii and Makarova, Anna and Skalnaya, Anastasia and Arkhipova, Nastasia and Sinkin, Mikhail and Ossadtchi, Alexei},
  journal={Journal of Neural Engineering},
  volume={19},
  number={6},
  pages={066016},
  year={2022},
  publisher={IOP Publishing}
}

@article{petrosyan2021decoding,
  title={Decoding and interpreting cortical signals with a compact convolutional neural network},
  author={Petrosyan, Artur and Sinkin, Mikhail and Lebedev, Mikhail and Ossadtchi, Alexei},
  journal={Journal of Neural Engineering},
  volume={18},
  number={2},
  pages={026019},
  year={2021},
  publisher={IOP Publishing}
}

@article{kumar2022neurophysiologically,
  title={A neurophysiologically interpretable deep neural network predicts complex movement components from brain activity},
  author={Kumar, Neelesh and Michmizos, Konstantinos P},
  journal={Scientific reports},
  volume={12},
  number={1},
  pages={1101},
  year={2022},
  publisher={Nature Publishing Group UK London}
}

@article{fu2022single,
  title={Single-trial motor imagery electroencephalogram intention recognition by optimal discriminant hyperplane and interpretable discriminative rectangle mixture model},
  author={Fu, Rongrong and Xu, Dong and Li, Weishuai and Shi, Peiming},
  journal={Cognitive Neurodynamics},
  volume={16},
  number={5},
  pages={1073--1085},
  year={2022},
  publisher={Springer}
}

@article{chen2022novel,
  title={A novel brain-computer interface based on audio-assisted visual evoked EEG and spatial-temporal attention CNN},
  author={Chen, Guijun and Zhang, Xueying and Zhang, Jing and Li, Fenglian and Duan, Shufei},
  journal={Frontiers in Neurorobotics},
  year={2022},
  publisher={Frontiers Research Foundation}
}

@article{lawhern2018eegnet,
  title={EEGNet: a compact convolutional neural network for EEG-based brain--computer interfaces},
  author={Lawhern, Vernon J and Solon, Amelia J and Waytowich, Nicholas R and Gordon, Stephen M and Hung, Chou P and Lance, Brent J},
  journal={Journal of neural engineering},
  volume={15},
  number={5},
  pages={056013},
  year={2018},
  publisher={iOP Publishing}
}

@article{zygierewicz2022decoding,
  title={Decoding working memory-related information from repeated psychophysiological EEG experiments using convolutional and contrastive neural networks},
  author={{\.Z}ygierewicz, Jaros{\l}aw and Janik, Romuald A and Podolak, Igor T and Drozd, Alan and Malinowska, Urszula and Poziomska, Martyna and Wojciechowski, Jakub and Ogniewski, Pawe{\l} and Niedbalski, Pawe{\l} and Terczynska, Iwona and others},
  journal={Journal of Neural Engineering},
  volume={19},
  number={4},
  pages={046053},
  year={2022},
  publisher={IOP Publishing}
}

@article{islam2022explainable,
  title={Explainable machine learning methods for classification of brain states during visual perception},
  author={Islam, Robiul and Andreev, Andrey V and Shusharina, Natalia N and Hramov, Alexander E},
  journal={Mathematics},
  volume={10},
  number={15},
  pages={2819},
  year={2022},
  publisher={MDPI}
}

@article{borra2022bayesian,
  title={A Bayesian-optimized design for an interpretable convolutional neural network to decode and analyze the P300 response in autism},
  author={Borra, Davide and Magosso, Elisa and Castelo-Branco, Miguel and Sim{\~o}es, Marco},
  journal={Journal of Neural Engineering},
  volume={19},
  number={4},
  pages={046010},
  year={2022},
  publisher={IOP Publishing}
}

@article{zhao2019learning,
  title={Learning joint space--time--frequency features for EEG decoding on small labeled data},
  author={Zhao, Dongye and Tang, Fengzhen and Si, Bailu and Feng, Xisheng},
  journal={Neural Networks},
  volume={114},
  pages={67--77},
  year={2019},
  publisher={Elsevier}
}

@article{ieracitano2021novel,
  title={A novel explainable machine learning approach for EEG-based brain-computer interface systems},
  author={Ieracitano, Cosimo and Mammone, Nadia and Hussain, Amir and Morabito, Francesco Carlo},
  journal={Neural Computing and Applications},
  pages={1--14},
  year={2021},
  publisher={Springer}
}

@article{bang2021spatio,
  title={Spatio-spectral feature representation for motor imagery classification using convolutional neural networks},
  author={Bang, Ji-Seon and Lee, Min-Ho and Fazli, Siamac and Guan, Cuntai and Lee, Seong-Whan},
  journal={IEEE Transactions on Neural Networks and Learning Systems},
  volume={33},
  number={7},
  pages={3038--3049},
  year={2021},
  publisher={IEEE}
}

@article{masse2022classification,
  title={Classification of Electrophysiological Signatures With Explainable Artificial Intelligence: The Case of Alarm Detection in Flight Simulator},
  author={Mass{\'e}, Eva and Bartheye, Olivier and Fabre, Ludovic},
  journal={Frontiers in Neuroinformatics},
  volume={16},
  year={2022},
  publisher={Frontiers Media SA}
}

@article{hammer2022interpretable,
  title={Interpretable functional specialization emerges in deep convolutional networks trained on brain signals},
  author={Hammer, Ji{\v{r}}i and Schirrmeister, RT and Hartmann, K and Marusic, P and Schulze-Bonhage, A and Ball, T},
  journal={Journal of Neural Engineering},
  volume={19},
  number={3},
  pages={036006},
  year={2022},
  publisher={IOP Publishing}
}

@article{ravindran2022decoding,
  title={Decoding neural activity preceding balance loss during standing with a lower-limb exoskeleton using an interpretable deep learning model},
  author={Ravindran, Akshay Sujatha and Malaya, Christopher A and John, Isaac and Francisco, Gerard E and Layne, Charles and Contreras-Vidal, Jose L},
  journal={Journal of neural engineering},
  volume={19},
  number={3},
  pages={036015},
  year={2022},
  publisher={IOP Publishing}
}

@article{du2022ienet,
  title={IENet: a robust convolutional neural network for EEG based brain-computer interfaces},
  author={Du, Yipeng and Liu, Jian},
  journal={Journal of Neural Engineering},
  volume={19},
  number={3},
  pages={036031},
  year={2022},
  publisher={IOP Publishing}
}

@article{banville2022robust,
  title={Robust learning from corrupted EEG with dynamic spatial filtering},
  author={Banville, Hubert and Wood, Sean UN and Aimone, Chris and Engemann, Denis-Alexander and Gramfort, Alexandre},
  journal={NeuroImage},
  volume={251},
  pages={118994},
  year={2022},
  publisher={Elsevier}
}

@article{tajmirriahi2022interpretable,
  title={An Interpretable Convolutional Neural Network for P300 Detection: Analysis of Time Frequency Features for Limited Data},
  author={Tajmirriahi, Mahnoosh and Amini, Zahra and Rabbani, Hossein and Kafieh, Rahele},
  journal={IEEE Sensors Journal},
  volume={22},
  number={9},
  pages={8685--8692},
  year={2022},
  publisher={IEEE}
}

@article{van2022classification,
  title={Classification of event-related potentials with regularized spatiotemporal LCMV beamforming},
  author={Van Den Kerchove, Arne and Libert, Arno and Wittevrongel, Benjamin and Van Hulle, Marc M},
  journal={Applied Sciences},
  volume={12},
  number={6},
  pages={2918},
  year={2022},
  publisher={MDPI}
}

@article{niu2022knowledge,
  title={Knowledge-driven feature component interpretable network for motor imagery classification},
  author={Niu, Xu and Lu, Na and Kang, Jianghong and Cui, Zhiyan},
  journal={Journal of Neural Engineering},
  volume={19},
  number={1},
  pages={016032},
  year={2022},
  publisher={IOP Publishing}
}

@inproceedings{bang2022interpretable,
  title={Interpretable Convolutional Neural Networks for Subject-Independent Motor Imagery Classification},
  author={Bang, Ji-Seon and Lee, Seong-Whan},
  booktitle={2022 10th International Winter Conference on Brain-Computer Interface (BCI)},
  pages={1--5},
  year={2022},
  organization={IEEE}
}

@article{adadi2018peeking,
  title={Peeking inside the black-box: a survey on explainable artificial intelligence (XAI)},
  author={Adadi, Amina and Berrada, Mohammed},
  journal={IEEE access},
  volume={6},
  pages={52138--52160},
  year={2018},
  publisher={IEEE}
}

@article{stuart2022interpretable,
  title={An Interpretable Deep Learning Model for Speech Activity Detection Using Electrocorticographic Signals},
  author={Stuart, Morgan and Lesaja, Srdjan and Shih, Jerry J and Schultz, Tanja and Manic, Milos and Krusienski, Dean J},
  journal={IEEE Transactions on Neural Systems and Rehabilitation Engineering},
  volume={30},
  pages={2783--2792},
  year={2022},
  publisher={IEEE}
}

@inproceedings{qu2022eeg4home,
  title={EEG4Home: A Human-In-The-Loop Machine Learning Model for EEG-Based BCI},
  author={Qu, Xiaodong and Hickey, Timothy J},
  booktitle={Augmented Cognition: 16th International Conference, AC 2022, Held as Part of the 24th HCI International Conference, HCII 2022, Virtual Event, June 26--July 1, 2022, Proceedings},
  pages={162--172},
  year={2022},
  organization={Springer}
}

@article{salami2022eeg,
  title={EEG-ITNet: An explainable inception temporal convolutional network for motor imagery classification},
  author={Salami, Abbas and Andreu-Perez, Javier and Gillmeister, Helge},
  journal={IEEE Access},
  volume={10},
  pages={36672--36685},
  year={2022},
  publisher={IEEE}
}

@article{mcdermott2021artifacts,
  title={Artifacts in EEG-based BCI therapies: friend or foe?},
  author={McDermott, Eric James and Raggam, Philipp and Kirsch, Sven and Belardinelli, Paolo and Ziemann, Ulf and Zrenner, Christoph},
  journal={Sensors},
  volume={22},
  number={1},
  pages={96},
  year={2021},
  publisher={MDPI}
}

@article{choi2021non,
  title={Non--human primate epidural ECoG analysis using explainable deep learning technology},
  author={Choi, Hoseok and Lim, Seokbeen and Min, Kyeongran and Ahn, Kyoung-ha and Lee, Kyoung-Min and Jang, Dong Pyo},
  journal={Journal of Neural Engineering},
  volume={18},
  number={6},
  pages={066022},
  year={2021},
  publisher={IOP Publishing}
}

@article{dong2021explainable,
  title={Explainable convolutional neural network to investigate age-related changes in multi-order functional connectivity},
  author={Dong, Sunghee and Jin, Yan and Bak, SuJin and Yoon, Bumchul and Jeong, Jichai},
  journal={Electronics},
  volume={10},
  number={23},
  pages={3020},
  year={2021},
  publisher={MDPI}
}

@article{aellen2021convolutional,
  title={Convolutional neural networks for decoding electroencephalography responses and visualizing trial by trial changes in discriminant features},
  author={Aellen, Florence M and G{\"o}ktepe-Kavis, Pinar and Apostolopoulos, Stefanos and Tzovara, Athina},
  journal={Journal of neuroscience methods},
  volume={364},
  pages={109367},
  year={2021},
  publisher={Elsevier}
}

@inproceedings{petrosyan2021compact,
  title={Compact and interpretable architecture for speech decoding from stereotactic EEG},
  author={Petrosyan, Artur and Voskoboynikov, Alexey and Ossadtchi, Alexei},
  booktitle={2021 Third International Conference Neurotechnologies and Neurointerfaces (CNN)},
  pages={79--82},
  year={2021},
  organization={IEEE}
}

@inproceedings{kobler2021interpretation,
  title={On the interpretation of linear Riemannian tangent space model parameters in M/EEG},
  author={Kobler, Reinmar J and Hirayama, Jun-Ichiro and Hehenberger, Lea and Lopes-Dias, Catarina and M{\"u}ller-Putz, Gernot R and Kawanabe, Motoaki},
  booktitle={2021 43rd Annual International Conference of the IEEE Engineering in Medicine \& Biology Society (EMBC)},
  pages={5909--5913},
  year={2021},
  organization={IEEE}
}

@article{lekova2021fuzzy,
  title={A fuzzy shell for developing an interpretable BCI based on the spatiotemporal dynamics of the evoked oscillations},
  author={Lekova, Anna and Chavdarov, Ivan},
  journal={Computational Intelligence and Neuroscience},
  volume={2021},
  year={2021},
  publisher={Hindawi}
}

@article{fu2021recognizing,
  title={Recognizing single-trial motor imagery EEG based on interpretable clustering method},
  author={Fu, Rongrong and Li, Weishuai and Chen, Junxiang and Han, Mengmeng},
  journal={Biomedical Signal Processing and Control},
  volume={63},
  pages={102171},
  year={2021},
  publisher={Elsevier}
}

@article{galindo2020multiple,
  title={Multiple kernel stein spatial patterns for the multiclass discrimination of motor imagery tasks},
  author={Galindo-Nore{\~n}a, Steven and C{\'a}rdenas-Pe{\~n}a, David and Orozco-Gutierrez, {\'A}lvaro},
  journal={Applied Sciences},
  volume={10},
  number={23},
  pages={8628},
  year={2020},
  publisher={MDPI}
}

@article{jin2020interpretable,
  title={Interpretable cross-subject EEG-based emotion recognition using channel-wise features},
  author={Jin, Longbin and Kim, Eun Yi},
  journal={Sensors},
  volume={20},
  number={23},
  pages={6719},
  year={2020},
  publisher={MDPI}
}

@article{borra2020interpretable,
  title={Interpretable and lightweight convolutional neural network for EEG decoding: Application to movement execution and imagination},
  author={Borra, Davide and Fantozzi, Silvia and Magosso, Elisa},
  journal={Neural Networks},
  volume={129},
  pages={55--74},
  year={2020},
  publisher={Elsevier}
}

@article{xu2020tangent,
  title={Tangent space spatial filters for interpretable and efficient Riemannian classification},
  author={Xu, Jiachen and Grosse-Wentrup, Moritz and Jayaram, Vinay},
  journal={Journal of Neural Engineering},
  volume={17},
  number={2},
  pages={026043},
  year={2020},
  publisher={IOP Publishing}
}

@article{huang2020spectrum,
  title={Spectrum-weighted tensor discriminant analysis for motor imagery-based BCI},
  author={Huang, Shoulin and Chen, Yang and Wang, Tong and Ma, Ting},
  journal={IEEE Access},
  volume={8},
  pages={93749--93759},
  year={2020},
  publisher={IEEE}
}

@article{loza2019discrimination,
  title={Discrimination of movement-related cortical potentials exploiting unsupervised learned representations from ECOGs},
  author={Loza, Carlos A and Reddy, Chandan G and Akella, Shailaja and Pr{\'\i}ncipe, Jos{\'e} C},
  journal={Frontiers in neuroscience},
  volume={13},
  pages={1248},
  year={2019},
  publisher={Frontiers Media SA}
}

@inproceedings{lopez2019supervised,
  title={Supervised Relevance Analysis for Multiple Stein Kernels for Spatio-Spectral Component Selection in BCI Discrimination Tasks},
  author={L{\'o}pez-Montes, Camilo and C{\'a}rdenas-Pe{\~n}a, David and Castellanos-Dominguez, Germ{\'a}n},
  booktitle={Progress in Pattern Recognition, Image Analysis, Computer Vision, and Applications: 24th Iberoamerican Congress, CIARP 2019, Havana, Cuba, October 28-31, 2019, Proceedings 24},
  pages={620--628},
  year={2019},
  organization={Springer}
}

@article{rahimi2018efficient,
  title={Efficient biosignal processing using hyperdimensional computing: Network templates for combined learning and classification of exg signals},
  author={Rahimi, Abbas and Kanerva, Pentti and Benini, Luca and Rabaey, Jan M},
  journal={Proceedings of the IEEE},
  volume={107},
  number={1},
  pages={123--143},
  year={2018},
  publisher={IEEE}
}

@article{motrenko2018multi,
  title={Multi-way feature selection for ECoG-based Brain-Computer Interface},
  author={Motrenko, Anastasia and Strijov, Vadim},
  journal={Expert Systems with Applications},
  volume={114},
  pages={402--413},
  year={2018},
  publisher={Elsevier}
}

@article{williams2018unsupervised,
  title={Unsupervised discovery of demixed, low-dimensional neural dynamics across multiple timescales through tensor component analysis},
  author={Williams, Alex H and Kim, Tony Hyun and Wang, Forea and Vyas, Saurabh and Ryu, Stephen I and Shenoy, Krishna V and Schnitzer, Mark and Kolda, Tamara G and Ganguli, Surya},
  journal={Neuron},
  volume={98},
  number={6},
  pages={1099--1115},
  year={2018},
  publisher={Elsevier}
}

@inproceedings{bouchard2017sparse,
  title={Sparse coding of ECoG signals identifies interpretable components for speech control in human sensorimotor cortex},
  author={Bouchard, Kristofer E and Bujan, Alejandro F and Chang, Edward F and Sommer, Friedrich T},
  booktitle={2017 39th Annual International Conference of the IEEE Engineering in Medicine and Biology Society (EMBC)},
  pages={3636--3639},
  year={2017},
  organization={IEEE}
}

@article{caywood2017gaussian,
  title={Gaussian process regression for predictive but interpretable machine learning models: an example of predicting mental workload across tasks},
  author={Caywood, Matthew S and Roberts, Daniel M and Colombe, Jeffrey B and Greenwald, Hal S and Weiland, Monica Z},
  journal={Frontiers in human neuroscience},
  volume={10},
  pages={647},
  year={2017},
  publisher={Frontiers Media SA}
}

@article{sturm2016interpretable,
  title={Interpretable deep neural networks for single-trial EEG classification},
  author={Sturm, Irene and Lapuschkin, Sebastian and Samek, Wojciech and M{\"u}ller, Klaus-Robert},
  journal={Journal of neuroscience methods},
  volume={274},
  pages={141--145},
  year={2016},
  publisher={Elsevier}
}

@article{wang2016unsupervised,
  title={Unsupervised decoding of long-term, naturalistic human neural recordings with automated video and audio annotations},
  author={Wang, Nancy XR and Olson, Jared D and Ojemann, Jeffrey G and Rao, Rajesh PN and Brunton, Bingni W},
  journal={Frontiers in human neuroscience},
  volume={10},
  pages={165},
  year={2016},
  publisher={Frontiers Media SA}
}

@article{bastos2016discovering,
  title={Discovering patterns in brain signals using decision trees},
  author={Bastos, Narusci S and Adamatti, Diana F and Billa, Cleo Z},
  journal={Computational Intelligence and Neuroscience},
  volume={2016},
  year={2016},
  publisher={Hindawi}
}

@article{dyson2015online,
  title={Online extraction and single trial analysis of regions contributing to erroneous feedback detection},
  author={Dyson, Matthew and Thomas, Eoin and Casini, Laurence and Burle, Boris},
  journal={NeuroImage},
  volume={121},
  pages={146--158},
  year={2015},
  publisher={Elsevier}
}

@inproceedings{mametkulov2022explainable,
  title={Explainable machine learning for memory-related decoding via TabNet and non-linear features*},
  author={Mametkulov, Maxim and Artykbayev, Abay and Koishigarina, Darina and Kenessova, Amina and Razikhova, Kamilla and Kang, Taeho and Wallraven, Christian and Fazli, Siamac},
  booktitle={2022 10th International Winter Conference on Brain-Computer Interface (BCI)},
  pages={1--7},
  year={2022},
  organization={IEEE}
}

@article{dong2023heterogeneous,
  title={Heterogeneous domain adaptation for intracortical signal classification using domain consensus},
  author={Dong, Yuanrui and Hu, Dingyin and Wang, Shirong and He, Jiping},
  journal={Biomedical Signal Processing and Control},
  volume={82},
  pages={104540},
  year={2023},
  publisher={Elsevier}
}

@article{cui2022compact,
  title={A compact and interpretable convolutional neural network for cross-subject driver drowsiness detection from single-channel EEG},
  author={Cui, Jian and Lan, Zirui and Liu, Yisi and Li, Ruilin and Li, Fan and Sourina, Olga and M{\"u}ller-Wittig, Wolfgang},
  journal={Methods},
  volume={202},
  pages={173--184},
  year={2022},
  publisher={Elsevier}
}

@inproceedings{zhao2023signal,
  title={Signal based Dilation Convolution CAM for Feature Extraction and Analysis in CNN Model},
  author={Zhao, Liang and Yang, Yifei and Feng, Yang and Xiao, Xundong and Wang, Xingjun},
  booktitle={Journal of Physics: Conference Series},
  volume={2425},
  pages={012010},
  year={2023},
  organization={IOP Publishing}
}

@article{kim2021deep,
  title={Deep-Learning-Based Automatic Selection of Fewest Channels for brain--machine Interfaces},
  author={Kim, Hyun-Seok and Ahn, Min-Hee and Min, Byoung-Kyong},
  journal={IEEE Transactions on Cybernetics},
  volume={52},
  number={9},
  pages={8668--8680},
  year={2021},
  publisher={IEEE}
}

@article{cui2022eeg,
  title={EEG-based cross-subject driver drowsiness recognition with an interpretable convolutional neural network},
  author={Cui, Jian and Lan, Zirui and Sourina, Olga and M{\"u}ller-Wittig, Wolfgang},
  journal={IEEE Transactions on Neural Networks and Learning Systems},
  year={2022},
  publisher={IEEE}
}

@phdthesis{kia2017brain,
  title={Brain Decoding for Brain Mapping: Definition, Heuristic Quantification, and Improvement of Interpretability in Group MEG Decoding},
  author={Kia, Seyed Mostafa},
  year={2017},
  school={University of Trento}
}

@article{lee2022quantifying,
  title={Quantifying arousal and awareness in altered states of consciousness using interpretable deep learning},
  author={Lee, Minji and Sanz, Leandro RD and Barra, Alice and Wolff, Audrey and Nieminen, Jaakko O and Boly, Melanie and Rosanova, Mario and Casarotto, Silvia and Bodart, Olivier and Annen, Jitka and others},
  journal={Nature communications},
  volume={13},
  number={1},
  pages={1064},
  year={2022},
  publisher={Nature Publishing Group UK London}
}

@article{kuang2023seer,
  title={SEER-net: Simple EEG-based Recognition network},
  author={Kuang, Dongyang and Michoski, Craig},
  journal={Biomedical Signal Processing and Control},
  volume={83},
  pages={104620},
  year={2023},
  publisher={Elsevier}
}

@article{thanigaivelu2023oisvm,
  title={OISVM: Optimal Incremental Support Vector Machine-based EEG Classification for Brain-computer Interface Model},
  author={Thanigaivelu, PS and Sridhar, SS and Sulthana, S Fouziya},
  journal={Cognitive Computation},
  pages={1--16},
  year={2023},
  publisher={Springer}
}

@article{tan2023eeg,
  title={EEG decoding for effects of visual joint attention training on ASD patients with interpretable and lightweight convolutional neural network},
  author={Tan, Jianling and Zhan, Yichao and Tang, Yi and Bao, Weixin and Tian, Yin},
  journal={Cognitive Neurodynamics},
  pages={1--14},
  year={2023},
  publisher={Springer}
}

@article{tanaka2020group,
  title={Group task-related component analysis (gTRCA): A multivariate method for inter-trial reproducibility and inter-subject similarity maximization for EEG data analysis},
  author={Tanaka, Hirokazu},
  journal={Scientific Reports},
  volume={10},
  number={1},
  pages={84},
  year={2020},
  publisher={Nature Publishing Group UK London}
}

@article{hu2020assessment,
  title={Assessment of nonnegative matrix factorization algorithms for electroencephalography spectral analysis},
  author={Hu, Guoqiang and Zhou, Tianyi and Luo, Siwen and Mahini, Reza and Xu, Jing and Chang, Yi and Cong, Fengyu},
  journal={BioMedical Engineering OnLine},
  volume={19},
  number={1},
  pages={1--18},
  year={2020},
  publisher={BioMed Central}
}

@article{kostas2019machine,
  title={Machine learning for MEG during speech tasks},
  author={Kostas, Demetres and Pang, Elizabeth W and Rudzicz, Frank},
  journal={Scientific reports},
  volume={9},
  number={1},
  pages={1--13},
  year={2019},
  publisher={Springer}
}

@article{collazos2020cnn,
  title={CNN-based framework using spatial dropping for enhanced interpretation of neural activity in motor imagery classification},
  author={Collazos-Huertas, Diego Fabian and {\'A}lvarez-Meza, Andr{\'e}s Marino and Acosta-Medina, Carlos Daniel and Casta{\~n}o-Duque, GA and Castellanos-Dominguez, Germ{\'a}n},
  journal={Brain Informatics},
  volume={7},
  number={1},
  pages={8},
  year={2020},
  publisher={Springer}
}

@article{vidaurre2023identification,
  title={Identification of spatial patterns with maximum association between power of resting state neural oscillations and trait anxiety},
  author={Vidaurre, Carmen and Nikulin, Vadim V and Herrojo Ruiz, Maria},
  journal={Neural Computing and Applications},
  volume={35},
  number={8},
  pages={5737--5749},
  year={2023},
  publisher={Springer}
}

@article{park2022individualized,
  title={Individualized diagnosis of preclinical Alzheimer’s Disease using deep neural networks},
  author={Park, Jinhee and Jang, Sehyeon and Gwak, Jeonghwan and Kim, Byeong C and Lee, Jang Jae and Choi, Kyu Yeong and Lee, Kun Ho and Jun, Sung Chan and Jang, Gil-Jin and Ahn, Sangtae},
  journal={Expert Systems with Applications},
  volume={210},
  pages={118511},
  year={2022},
  publisher={Elsevier}
}

@article{jiang2020smart,
  title={Smart diagnosis: a multiple-source transfer TSK fuzzy system for EEG seizure identification},
  author={Jiang, Yizhang and Gu, Xiaoqing and Ji, Dingcheng and Qian, Pengjiang and Xue, Jing and Zhang, Yuanpeng and Zhu, Jiaqi and Xia, Kaijian and Wang, Shitong},
  journal={ACM Transactions on Multimedia Computing, Communications, and Applications (TOMM)},
  volume={16},
  number={2s},
  pages={1--21},
  year={2020},
  publisher={ACM New York, NY, USA}
}

@article{raab2022xai4eeg,
  title={XAI4EEG: spectral and spatio-temporal explanation of deep learning-based seizure detection in EEG time series},
  author={Raab, Dominik and Theissler, Andreas and Spiliopoulou, Myra},
  journal={Neural Computing and Applications},
  pages={1--18},
  year={2022},
  publisher={Springer}
}

@article{petrescu2021machine,
  title={Machine learning methods for fear classification based on physiological features},
  author={Petrescu, Livia and Petrescu, C{\u{a}}t{\u{a}}lin and Oprea, Ana and Mitru , Oana and Moise, Gabriela and Moldoveanu, Alin and Moldoveanu, Florica},
  journal={Sensors},
  volume={21},
  number={13},
  pages={4519},
  year={2021},
  publisher={MDPI}
}

@article{svetlakov2022representation,
  title={Representation Learning for EEG-Based Biometrics Using Hilbert--Huang Transform},
  author={Svetlakov, Mikhail and Kovalev, Ilya and Konev, Anton and Kostyuchenko, Evgeny and Mitsel, Artur},
  journal={Computers},
  volume={11},
  number={3},
  pages={47},
  year={2022},
  publisher={MDPI}
}

@article{lo2022permutation,
  title={Permutation entropy-based interpretability of convolutional neural network models for interictal eeg discrimination of subjects with epileptic seizures vs. psychogenic non-epileptic seizures},
  author={Lo Giudice, Michele and Varone, Giuseppe and Ieracitano, Cosimo and Mammone, Nadia and Tripodi, Giovanbattista Gaspare and Ferlazzo, Edoardo and Gasparini, Sara and Aguglia, Umberto and Morabito, Francesco Carlo},
  journal={Entropy},
  volume={24},
  number={1},
  pages={102},
  year={2022},
  publisher={MDPI}
}

@article{na2022objective,
  title={Objective speech intelligibility prediction using a deep learning model with continuous speech-evoked cortical auditory responses},
  author={Na, Youngmin and Joo, Hyosung and Trang, Le Thi and Quan, Luong Do Anh and Woo, Jihwan},
  journal={Frontiers in Neuroscience},
  pages={1352},
  year={2022},
  publisher={Frontiers}
}

@inproceedings{zhang2020eeg,
  title={EEG-Based Short-Time Auditory Attention Detection Using Multi-Task Deep Learning.},
  author={Zhang, Zhuo and Zhang, Gaoyan and Dang, Jianwu and Wu, Shuang and Zhou, Di and Wang, Longbiao},
  booktitle={INTERSPEECH},
  pages={2517--2521},
  year={2020}
}

@inproceedings{hsieh2021explainable,
  title={Explainable multivariate time series classification: a deep neural network which learns to attend to important variables as well as time intervals},
  author={Hsieh, Tsung-Yu and Wang, Suhang and Sun, Yiwei and Honavar, Vasant},
  booktitle={Proceedings of the 14th ACM international conference on web search and data mining},
  pages={607--615},
  year={2021}
}

@article{gabeff2021interpreting,
  title={Interpreting deep learning models for epileptic seizure detection on EEG signals},
  author={Gabeff, Valentin and Teijeiro, Tomas and Zapater, Marina and Cammoun, Leila and Rheims, Sylvain and Ryvlin, Philippe and Atienza, David},
  journal={Artificial intelligence in medicine},
  volume={117},
  pages={102084},
  year={2021},
  publisher={Elsevier}
}

@inproceedings{dhanorkar2021needs,
  title={Who needs to know what, when?: Broadening the Explainable AI (XAI) Design Space by Looking at Explanations Across the AI Lifecycle},
  author={Dhanorkar, Shipi and Wolf, Christine T and Qian, Kun and Xu, Anbang and Popa, Lucian and Li, Yunyao},
  booktitle={Designing Interactive Systems Conference 2021},
  pages={1591--1602},
  year={2021}
}

@article{saeed2023explainable,
  title={Explainable ai (xai): A systematic meta-survey of current challenges and future opportunities},
  author={Saeed, Waddah and Omlin, Christian},
  journal={Knowledge-Based Systems},
  pages={110273},
  year={2023},
  publisher={Elsevier}
}

@article{moore2003real,
  title={Real-world applications for brain-computer interface technology},
  author={Moore, Melody M},
  journal={IEEE Transactions on Neural Systems and Rehabilitation Engineering},
  volume={11},
  number={2},
  pages={162--165},
  year={2003},
  publisher={IEEE}
}

@article{moore2010applications,
  title={Applications for brain-computer interfaces},
  author={Moore Jackson, Melody and Mappus, Rudolph},
  journal={Brain-computer interfaces: applying our minds to human-computer interaction},
  pages={89--103},
  year={2010},
  publisher={Springer}
}

@article{mak2009clinical,
  title={Clinical applications of brain-computer interfaces: current state and future prospects},
  author={Mak, Joseph N and Wolpaw, Jonathan R},
  journal={IEEE reviews in biomedical engineering},
  volume={2},
  pages={187--199},
  year={2009},
  publisher={ieee}
}

@article{yin2020locally,
  title={Locally robust EEG feature selection for individual-independent emotion recognition},
  author={Yin, Zhong and Liu, Lei and Chen, Jianing and Zhao, Boxi and Wang, Yongxiong},
  journal={Expert Systems with Applications},
  volume={162},
  pages={113768},
  year={2020},
  publisher={Elsevier}
}

@inproceedings{selvaraju2017grad,
  title={Grad-cam: Visual explanations from deep networks via gradient-based localization},
  author={Selvaraju, Ramprasaath R and Cogswell, Michael and Das, Abhishek and Vedantam, Ramakrishna and Parikh, Devi and Batra, Dhruv},
  booktitle={Proceedings of the IEEE international conference on computer vision},
  pages={618--626},
  year={2017}
}

@inproceedings{karimi2021algorithmic,
  title={Algorithmic recourse: from counterfactual explanations to interventions},
  author={Karimi, Amir-Hossein and Sch{\"o}lkopf, Bernhard and Valera, Isabel},
  booktitle={Proceedings of the 2021 ACM conference on fairness, accountability, and transparency},
  pages={353--362},
  year={2021}
}

@inproceedings{howard2020we,
  title={Are we trusting AI too much? Examining human-robot interactions in the real world},
  author={Howard, Ayanna},
  booktitle={Proceedings of the 2020 ACM/IEEE International Conference on Human-Robot Interaction},
  pages={1--1},
  year={2020}
}

@article{ghassemi2021false,
  title={The false hope of current approaches to explainable artificial intelligence in health care},
  author={Ghassemi, Marzyeh and Oakden-Rayner, Luke and Beam, Andrew L},
  journal={The Lancet Digital Health},
  volume={3},
  number={11},
  pages={e745--e750},
  year={2021},
  publisher={Elsevier}
}

@article{markus2021role,
  title={The role of explainability in creating trustworthy artificial intelligence for health care: a comprehensive survey of the terminology, design choices, and evaluation strategies},
  author={Markus, Aniek F and Kors, Jan A and Rijnbeek, Peter R},
  journal={Journal of Biomedical Informatics},
  volume={113},
  pages={103655},
  year={2021},
  publisher={Elsevier}
}

@article{nauta2022anecdotal,
  title={From anecdotal evidence to quantitative evaluation methods: A systematic review on evaluating explainable ai},
  author={Nauta, Meike and Trienes, Jan and Pathak, Shreyasi and Nguyen, Elisa and Peters, Michelle and Schmitt, Yasmin and Schl{\"o}tterer, J{\"o}rg and van Keulen, Maurice and Seifert, Christin},
  journal={arXiv preprint arXiv:2201.08164},
  year={2022}
}

@article{apicella2022toward,
  title={Toward the application of XAI methods in EEG-based systems},
  author={Apicella, Andrea and Isgr{\`o}, Francesco and Pollastro, Andrea and Prevete, Roberto},
  journal={arXiv preprint arXiv:2210.06554},
  year={2022}
}

@inproceedings{speith2022review,
  title={A review of taxonomies of explainable artificial intelligence (XAI) methods},
  author={Speith, Timo},
  booktitle={2022 ACM Conference on Fairness, Accountability, and Transparency},
  pages={2239--2250},
  year={2022}
}

@article{angelov2021explainable,
  title={Explainable artificial intelligence: an analytical review},
  author={Angelov, Plamen P and Soares, Eduardo A and Jiang, Richard and Arnold, Nicholas I and Atkinson, Peter M},
  journal={Wiley Interdisciplinary Reviews: Data Mining and Knowledge Discovery},
  volume={11},
  number={5},
  pages={e1424},
  year={2021},
  publisher={Wiley Online Library}
}

@article{arya2019one,
  title={One explanation does not fit all: A toolkit and taxonomy of ai explainability techniques},
  author={Arya, Vijay and Bellamy, Rachel KE and Chen, Pin-Yu and Dhurandhar, Amit and Hind, Michael and Hoffman, Samuel C and Houde, Stephanie and Liao, Q Vera and Luss, Ronny and Mojsilovi{\'c}, Aleksandra and others},
  journal={arXiv preprint arXiv:1909.03012},
  year={2019}
}

@incollection{stankiewicz2000concept,
  title={The concept of “design space”},
  author={Stankiewicz, Rikard},
  booktitle={Technological innovation as an evolutionary process},
  pages={234--247},
  year={2000}
}

@article{shneiderman2020human,
  title={Human-centered artificial intelligence: Reliable, safe \& trustworthy},
  author={Shneiderman, Ben},
  journal={International Journal of Human--Computer Interaction},
  volume={36},
  number={6},
  pages={495--504},
  year={2020},
  publisher={Taylor \& Francis}
}

@article{kosmyna2019conceptual,
  title={A conceptual space for EEG-based brain-computer interfaces},
  author={Kosmyna, Nataliya and L{\'e}cuyer, Anatole},
  journal={PloS one},
  volume={14},
  number={1},
  pages={e0210145},
  year={2019},
  publisher={Public Library of Science San Francisco, CA USA}
}

@article{mason2003general,
  title={A general framework for brain-computer interface design},
  author={Mason, Steven G and Birch, Gary E},
  journal={IEEE transactions on neural systems and rehabilitation engineering},
  volume={11},
  number={1},
  pages={70--85},
  year={2003},
  publisher={IEEE}
}

@article{lundberg2017unified,
  title={A unified approach to interpreting model predictions},
  author={Lundberg, Scott M and Lee, Su-In},
  journal={Advances in neural information processing systems},
  volume={30},
  year={2017}
}

@inproceedings{shrikumar2017learning,
  title={Learning important features through propagating activation differences},
  author={Shrikumar, Avanti and Greenside, Peyton and Kundaje, Anshul},
  booktitle={International conference on machine learning},
  pages={3145--3153},
  year={2017},
  organization={PMLR}
}

@article{bach2015pixel,
  title={On pixel-wise explanations for non-linear classifier decisions by layer-wise relevance propagation},
  author={Bach, Sebastian and Binder, Alexander and Montavon, Gr{\'e}goire and Klauschen, Frederick and M{\"u}ller, Klaus-Robert and Samek, Wojciech},
  journal={PloS one},
  volume={10},
  number={7},
  pages={e0130140},
  year={2015},
  publisher={Public Library of Science}
}

@inproceedings{sundararajan2017axiomatic,
  title={Axiomatic attribution for deep networks},
  author={Sundararajan, Mukund and Taly, Ankur and Yan, Qiqi},
  booktitle={International conference on machine learning},
  pages={3319--3328},
  year={2017},
  organization={PMLR}
}

@article{simonyan2013deep,
  title={Deep inside convolutional networks: Visualising image classification models and saliency maps},
  author={Simonyan, Karen and Vedaldi, Andrea and Zisserman, Andrew},
  journal={arXiv preprint arXiv:1312.6034},
  year={2013}
}

@article{van2008visualizing,
  title={Visualizing data using t-SNE.},
  author={Van der Maaten, Laurens and Hinton, Geoffrey},
  journal={Journal of machine learning research},
  volume={9},
  number={11},
  year={2008}
}

@article{ahn2012feasibility,
  title={Feasibility of approaches combining sensor and source features in brain--computer interface},
  author={Ahn, Minkyu and Hong, Jun Hee and Jun, Sung Chan},
  journal={Journal of neuroscience methods},
  volume={204},
  number={1},
  pages={168--178},
  year={2012},
  publisher={Elsevier}
}

@article{zhang2021tiny,
  title={Tiny noise, big mistakes: adversarial perturbations induce errors in brain--computer interface spellers},
  author={Zhang, Xiao and Wu, Dongrui and Ding, Lieyun and Luo, Hanbin and Lin, Chin-Teng and Jung, Tzyy-Ping and Chavarriaga, Ricardo},
  journal={National Science Review},
  volume={8},
  number={4},
  pages={nwaa233},
  year={2021},
  publisher={Oxford University Press}
}

@article{meng2023adversarial,
  title={Adversarial robustness benchmark for EEG-based brain--computer interfaces},
  author={Meng, Lubin and Jiang, Xue and Wu, Dongrui},
  journal={Future Generation Computer Systems},
  volume={143},
  pages={231--247},
  year={2023},
  publisher={Elsevier}
}

@article{etmann2019connection,
  title={On the connection between adversarial robustness and saliency map interpretability},
  author={Etmann, Christian and Lunz, Sebastian and Maass, Peter and Sch{\"o}nlieb, Carola-Bibiane},
  journal={arXiv preprint arXiv:1905.04172},
  year={2019}
}

@InProceedings{Chan_2020_CVPR,
author = {Chan, Alvin and Tay, Yi and Ong, Yew-Soon},
title = {What It Thinks Is Important Is Important: Robustness Transfers Through Input Gradients},
booktitle = {Proceedings of the IEEE/CVF Conference on Computer Vision and Pattern Recognition (CVPR)},
month = {June},
year = {2020}
}

@inproceedings{ross2018improving,
  title={Improving the adversarial robustness and interpretability of deep neural networks by regularizing their input gradients},
  author={Ross, Andrew and Doshi-Velez, Finale},
  booktitle={Proceedings of the AAAI conference on artificial intelligence},
  volume={32},
  number={1},
  year={2018}
}

@article{zhou2023interpretable,
  title={Interpretable and robust ai in eeg systems: A survey},
  author={Zhou, Xinliang and Liu, Chenyu and Zhai, Liming and Jia, Ziyu and Guan, Cuntai and Liu, Yang},
  journal={arXiv preprint arXiv:2304.10755},
  year={2023}
}

@article{weber2022beyond,
  title={Beyond explaining: Opportunities and challenges of XAI-based model improvement},
  author={Weber, Leander and Lapuschkin, Sebastian and Binder, Alexander and Samek, Wojciech},
  journal={Information Fusion},
  year={2022},
  publisher={Elsevier}
}

@inproceedings{kumar2020problems,
  title={Problems with Shapley-value-based explanations as feature importance measures},
  author={Kumar, I Elizabeth and Venkatasubramanian, Suresh and Scheidegger, Carlos and Friedler, Sorelle},
  booktitle={International Conference on Machine Learning},
  pages={5491--5500},
  year={2020},
  organization={PMLR}
}

@article{verdinelli2023feature,
  title={Feature Importance: A Closer Look at Shapley Values and LOCO},
  author={Verdinelli, Isabella and Wasserman, Larry},
  journal={arXiv preprint arXiv:2303.05981},
  year={2023}
}

@article{ehsan2021explainability,
  title={Explainability pitfalls: Beyond dark patterns in explainable AI},
  author={Ehsan, Upol and Riedl, Mark O},
  journal={arXiv preprint arXiv:2109.12480},
  year={2021}
}

@inproceedings{chromik2019dark,
  title={Dark Patterns of Explainability, Transparency, and User Control for Intelligent Systems.},
  author={Chromik, Michael and Eiband, Malin and V{\"o}lkel, Sarah Theres and Buschek, Daniel},
  booktitle={IUI workshops},
  volume={2327},
  year={2019}
}

@inproceedings{anders2020fairwashing,
  title={Fairwashing explanations with off-manifold detergent},
  author={Anders, Christopher and Pasliev, Plamen and Dombrowski, Ann-Kathrin and M{\"u}ller, Klaus-Robert and Kessel, Pan},
  booktitle={International Conference on Machine Learning},
  pages={314--323},
  year={2020},
  organization={PMLR}
}

@article{dombrowski2019explanations,
  title={Explanations can be manipulated and geometry is to blame},
  author={Dombrowski, Ann-Kathrin and Alber, Maximillian and Anders, Christopher and Ackermann, Marcel and M{\"u}ller, Klaus-Robert and Kessel, Pan},
  journal={Advances in neural information processing systems},
  volume={32},
  year={2019}
}

@article{han2022explanation,
  title={Which explanation should i choose? a function approximation perspective to characterizing post hoc explanations},
  author={Han, Tessa and Srinivas, Suraj and Lakkaraju, Himabindu},
  journal={Advances in Neural Information Processing Systems},
  volume={35},
  pages={5256--5268},
  year={2022}
}

@article{reddy2022explainability,
  title={Explainability and artificial intelligence in medicine},
  author={Reddy, Sandeep},
  journal={The Lancet Digital Health},
  volume={4},
  number={4},
  pages={e214--e215},
  year={2022},
  publisher={Elsevier}
}

@article{gwon2023review,
  title={Review of public motor imagery and execution datasets in brain-computer interfaces},
  author={Gwon, Daeun and Won, Kyungho and Song, Minseok and Nam, Chang S and Jun, Sung Chan and Ahn, Minkyu},
  journal={Frontiers in Human Neuroscience},
  volume={17},
  pages={1134869},
  year={2023},
  publisher={Frontiers}
}

@article{warrens2015five,
  title={Five ways to look at Cohen's kappa},
  author={Warrens, Matthijs J},
  journal={Journal of Psychology \& Psychotherapy},
  volume={5},
  year={2015},
  publisher={OMICS International}
}

@inproceedings{sokolova2006beyond,
  title={Beyond accuracy, F-score and ROC: a family of discriminant measures for performance evaluation},
  author={Sokolova, Marina and Japkowicz, Nathalie and Szpakowicz, Stan},
  booktitle={Australasian joint conference on artificial intelligence},
  pages={1015--1021},
  year={2006},
  organization={Springer}
}

@article{haufe2014interpretation,
  title={On the interpretation of weight vectors of linear models in multivariate neuroimaging},
  author={Haufe, Stefan and Meinecke, Frank and G{\"o}rgen, Kai and D{\"a}hne, Sven and Haynes, John-Dylan and Blankertz, Benjamin and Bie{\ss}mann, Felix},
  journal={Neuroimage},
  volume={87},
  pages={96--110},
  year={2014},
  publisher={Elsevier}
}

@article{montavon2019layer,
  title={Layer-wise relevance propagation: an overview},
  author={Montavon, Gr{\'e}goire and Binder, Alexander and Lapuschkin, Sebastian and Samek, Wojciech and M{\"u}ller, Klaus-Robert},
  journal={Explainable AI: interpreting, explaining and visualizing deep learning},
  pages={193--209},
  year={2019},
  publisher={Springer}
}

@article{niu2021review,
  title={A review on the attention mechanism of deep learning},
  author={Niu, Zhaoyang and Zhong, Guoqiang and Yu, Hui},
  journal={Neurocomputing},
  volume={452},
  pages={48--62},
  year={2021},
  publisher={Elsevier}
}

% \printbibliography

\end{document}